\documentclass[twocolumn,epjc3]{svjour3}
\usepackage{graphicx}
\usepackage{epsfig}
\usepackage{amsmath,amssymb}
\usepackage{color}
\journalname{Eur. Phys. J. C}
\begin{document}
%\documentclass[amsmath,amssymb,twocolumn,nofootinbib]{revtex4-1}
%\usepackage{graphicx}
%\usepackage{epsfig}
%\begin{document}
\title{Massive scalar field in de Sitter spacetime:\\a two-loop calculation and a comparison with the stochastic approach}
\author{Alexander~Yu.~Kamenshchik\thanksref{e1,addr1,addr2},
Alexei~A.~Starobinsky\thanksref{e2,addr2,addr3}
\and
Tereza~Vardanyan\thanksref{e3,addr4}}
\thankstext{e1}{e-mail: kamenshchik@bo.infn.it}
\thankstext{e2}{e-mail: alstar@landau.ac.ru}
\thankstext{e3}{e-mail: tereza.vardanyan@univaq.it}
\institute{Dipartimento di Fisica e Astronomia, Universit\`a di Bologna and INFN,
via Irnerio 46, 40126 Bologna, Italy \label{addr1}
\and
L. D. Landau Institute for
Theoretical Physics, Moscow, 119334 Russia\label{addr2}
\and
Bogolyubov Laboratory of Theoretical Physics, Joint Institute for Nuclear Research, 141980 Dubna, Russia
\label{addr3}
\and
Dipartimento di Fisica e Chimica, Universit\`a di L'Aquila, 67100 Coppito, L'Aquila, Italy\\ and 
 INFN, Laboratori Nazionali del Gran Sasso, 67010 Assergi, L'Aquila, Italy\label{addr4}
}
%\author{Alexander~Yu.~Kamenshchik}
%\email{kamenshchik@bo.infn.it}
%\affiliation{Dipartimento di Fisica e Astronomia, Universit\`a di Bologna\\ and INFN,  Via Irnerio 46, 40126 Bologna,
%Italy,\\
%L.D. Landau Institute for Theoretical Physics of the Russian
%Academy of Sciences,
%119334 Moscow, Russia}
%\author{Alexei A. Starobinsky}
%\email{alstar@landau.ac.ru}
%\affiliation{L.D. Landau Institute for Theoretical Physics of the Russian Academy of Sciences, 119334 Moscow, Russia, \\
%Bogolyubov Laboratory of Theoretical Physics, Joint Institute for Nuclear Research, 141980 Dubna, Russia}
%\author{Tereza Vardanyan}
%\email{tereza.vardanyan@aquila.infn.it}
%\affiliation{Dipartimento di Fisica e Chimica, Universit\`a di L'Aquila, 67100 Coppito, L'Aquila, Italy\\ and 
%INFN, Laboratori Nazionali del Gran Sasso, 67010 Assergi, L'Aquila, Italy}
\date{Received: date / Accepted: date}
\maketitle
\begin{abstract}
We examine long-wavelength correlation functions of massive scalar fields in de Sitter spacetime. For the theory with a quartic self-interaction, the two-point function is calculated up to two loops. Comparing our results with the Hartree-Fock approximation and with the stochastic approach shows that the former resums only the cactus type diagrams, whereas the latter contains the sunset diagram as well and produces the correct result. We also demonstrate that the long-wavelength expectation value of the commutator of two fields is equal to zero both for spacelike and timelike separated points.      
\end{abstract}
\vspace{1cm}
\section{Introduction}
Quantum field theory in expanding spacetimes is crucial for exploring the origin and the evolution of the universe. Scalar fields in de Sitter spacetime are of particular importance for understanding the period of inflation and the growth of quantum fluctuations during this period. 

Free scalar fields on the de Sitter background---both massive and massless---have been studied extensively (see, e.g., \cite{ChT,Tagirov,BD,Linde:1982uu,Starobinsky:1982ee,Vilenkin:1982wt,Allen:1985ux,Allen:1987tz}). It has been shown that for massive fields there exists a one-parameter family of de Sitter-invariant states; this includes the Bunch-Davies vacuum, whose modes in the Poincare patch match with the adiabatic in-vacuum modes in the remote past. From this family, only the Bunch-Davies vacuum has finite renormalized values of $\langle\phi^2\rangle$ and the energy-momentum tensor of the scalar field.

The massless minimally coupled scalar field also has the Bunch-Davies state, but it is no longer de Sitter invariant. It is, however, invariant under the $\rm E(3)$ subgroup, which is the group of translations and rotations in the flat spatial sections of de Sitter spacetime. One of the manifestations of the breakdown of de Sitter invariance is the presence of an infrared divergence in the theory. Hence, to evaluate correlation functions, an infrared cutoff should be introduced in momentum space. As a result, expectation values of products of fields are explicitly time-dependent; for example,
\begin{equation*}
\langle\phi^2(\vec x,t)\rangle_{\rm ren}={H^3t\over4\pi^2}+{\rm Constant}\, ,
\end{equation*}   
where $H$ is the constant Hubble parameter in the Poincare patch of de Sitter space (see \eqref{a} below) and $t=0$ denotes the beginning of de Sitter stage~\cite{Linde:1982uu,Starobinsky:1982ee,Vilenkin:1982wt} (see also~\cite{Allen:1985ux}), e.g., arising from a previous generic anisotropic and inhomogeneous stage with curvature much exceeding $H^2$ that does not require fine tuning of pre-inflationary initial conditions. At late times, the first term on the right side dominates. This time-dependent term arises solely from long-wavelength modes, that is, modes with physical wavelengths much greater than the Hubble scale. 

A similar result is obtained for small gauge-invariant metric tensor perturbations--- gravitational waves (GW)---of the de Sitter background~\cite{Starobinsky:1979ty}:
\begin{equation*}
\langle h_{ij}h^{ij}\rangle_{\rm ren}= \frac{16GH^3t}{\pi} + {\rm Constant}\, ,
\end{equation*} 
where $G$ is Newton's gravitational constant and $i,j=1,2,3$ (in~\cite{Starobinsky:1979ty} this is calculated in the synchronous gauge, but it does not matter, since the tensor perturbations $h_{ij}$ are gauge invariant). In the context of the inflationary scenario where the duration of de Sitter (inflationary) stage is finite, i.e. our past light cone is finite, this formula just corresponds to the standard expression for the primordial power spectrum of tensor perturbations generated during inflation, summed over polarizations: $P_g=\frac{16GH^2}{\pi}$, where slowly varying $H$ is estimated at the moment of the first Hubble radius crossing, $k/a(t)=H(t)$, for each mode with a comoving momentum k. In~\cite{Starobinsky:1979ty}, these results were explicitly presented in another equivalent form---as the energy density of primordial GW after the second Hubble radius crossing at the subsequent radiation dominated stage.

In the massive case there is no infrared divergence, but if $0<m^2\ll H^2$, the leading contribution to $\langle\phi^2\rangle_{\rm ren}$, which is given by \cite{Linde:1982uu,Starobinsky:1982ee,Vilenkin:1982wt,Allen:1987tz}
\begin{equation*}
\langle\phi^2(\vec x,t)\rangle_{\rm ren}={3H^4\over8\pi^2m^2}+\mathcal{O}\left(\left(m^2/H^2\right)^0\right),
\end{equation*}
still derives entirely from the long-wavelength modes.     

When there is a self-interaction, each successive term in the weak coupling perturbative expansion contains higher and higher powers of $Ht$ or $H^2/m^2$, depending on whether the theory is massless or massive. This means that perturbation theory breaks down when the value of $Ht$ or $H^2/m^2$ overwhelms the smallness of the coupling constant \cite{Woodard,Prokopec:2007ak,we}. A non-perturbative method for calculating the expectation values of the coarse-grained theory, containing only the long-wavelength fluctuations of a scalar field, was developed by Starobinsky in \cite{Star} and further developed in many papers, in particular in \cite{Star-Yoko,Vennin:2015hra}. An important result of Starobinsky's approach is that expectation values can be determined by using a probability distribution function that is a solution to a simple Fokker-Planck equation. The emergence of the Fokker-Planck equation from the full quantum evolution of the theory was presented in more recent works \cite{Tereza,Burgess:2014eoa,Burgess:2015ajz,Tokuda:2017fdh,Pinol:2020cdp,Baum-Sun,Gorb-Sen}.  

Due to the non-perturbative nature of Starobinsky's stochastic approach, it is important to check it by comparing its results with those obtained from the standard field-theoretic perturbative approach whenever the latter is possible. This was done in the one-loop (Hartree-Fock, or Gaussian) approximation in~\cite{Finelli:2008zg,Finelli:2010sh}. Here we consider the two-loop case in the regime when the scalar field mass and its quartic interaction are sufficiently small. A number of two-loop calculations for  scalar fields with quartic self-interaction in de Sitter spacetime were made  using the stochastic approach and  different quantum field theoretical methods  ~\cite{Marolf:2010nz,Gat-Ser,Hollands:2010pr,Akhmedov:2013xka,Garb-Rig-Zhu,Gautier:2015pca,Onemli,Onemli1,Mark-et,Kitamoto,Mor-Ser,Mark-Raj}. 

In our previous paper \cite{we}, we considered a massless minimally coupled scalar field with a quartic self-interaction. We calculated the long-wavelength part of $\langle\phi^2(\vec x,t)\rangle$ and $\langle\phi^4(\vec x,t)\rangle$ to second order in the coupling constant and developed a method for taking the late-time limit of these perturbative series. This method was inspired by the renormalization group and is based on the assumption that expectation values of products of fields at coinciding spacetime points satisfy autonomous first-order differential equations. We then compared our results with the results of the stochastic approach and found them to be in good agreement.    

In this paper we turn to the study of the massive case. In Sec.~2 we review the quantization of a free massive scalar field in de Sitter space. Section~3 presents a perturbative calculation of the long-wavelength part of the two-point function. Section~3.1 contains the zeroth-order computation and a comparison with the untruncated two-point function of the free theory. In Sec.~3.2 we outline the ``in-in'' formalism and use it to calculate the one- and two-loop corrections in Sections~3.3 and 3.4. For coinciding spacetime points our perturbative expression agrees with those obtained in \cite{Gat-Ser,Garb-Rig-Zhu,Mor-Ser}.
We compare our results with the Hartree-Fock approximation in Sec.~4  and with the stochastic approach in Sec.~5. In Sec.~6 we argue that the perturbative expression for the two-point function can be reorganized into a sum of exponential functions that depend on the two given points in a de Sitter invariant way. \\ 
\vspace{0.5cm}
\section{Scalar field in de Sitter spacetime} 
We will study a massive scalar field with a quartic self-interaction, 
\begin{equation}
S = \int d^4x \sqrt{-g}\left(\frac12 g^{\mu\nu}\partial_{\mu}\phi\partial_{\nu}\phi-{1\over2}{m^2}\phi^2-{\lambda\over4}\phi^4\right)\;,
\label{action1}
\end{equation}
propagating in the de Sitter background represented as an expanding spatially flat Friedmann universe (the Poincare patch) 
with the metric  
\begin{equation}
ds^2=dt^2-a^2(t)\delta_{ij}dx^idx^j\;,
\label{dS}
\end{equation}
where the scale factor $a(t)$ is 
\begin{equation}
a(t) = e^{Ht}\;.
\label{a}
\end{equation}
Here $t$ is a cosmic time coordinate and $H$ is the Hubble constant or the inverse of the de Sitter radius. The cosmic time in an expanding de Sitter universe runs in the interval $-\infty < t < \infty$. It will also be convenient to use a conformal time coordinate $\eta$, which is related to the cosmic time $t$ by the condition $dt=a(\eta)d\eta$. Expressed in terms of the conformal time, the metric is 
\begin{equation}
ds^2 = a^2(\eta)(d\eta^2-\delta_{ij}dx^idx^j)\;,
\label{conformal}
\end{equation}
where 
\begin{equation}
a(\eta) = -\frac{1}{H\eta}\;,
\label{conformal1}
\end{equation}
and $\eta$ runs from $-\infty$ to $0$. 

We will assume that the coupling constant and the mass are small: $\lambda\ll 1$, $m^2\ll H^2$. However, as we will see from the perturbative loop expansion, it has to be assumed additionally that the mass is not too small: 
$\sqrt{\lambda}H^2\ll m^2 \ll H^2$. Note that there is no need in the last condition for the stochastic approach, but we have to impose it since we want to compare the stochastic approach with the results of perturbation theory. 

When $\lambda=0$, the equation of motion for the rescaled field $\chi\equiv a(\eta)\phi$ is 
\begin{equation*}
\chi''-\nabla^2\chi-{1\over\eta^2}\left(2-{m^2\over H^2}\right)\chi=0\;,
\end{equation*}
where primes denote derivatives with respect to the conformal time and $\nabla^2$ is the three-dimensional Laplacian. 

To quantize the field $\phi(\vec{x},t)$, we expand it in terms of creation and annihilation operators  
\begin{equation*}
\phi(\vec{x},t) = \int {d^3\vec{k}\over (2\pi)^3}\left\{\phi_k(\eta)e^{i\vec{k}\cdot\vec{x}}a_{\vec{k}}+\phi_k^*(\eta)e^{-i\vec{k}\cdot\vec{x}}a_{\vec k}^\dagger\right\}\;,
\end{equation*}
where $a_{\vec k}^\dagger$ and $a_{\vec{k}}$ satisfy the usual commutation relations. The mode functions $\chi_k\equiv a(\eta)\phi_k$ then obey the differential equation
\begin{eqnarray*}
\chi_k''+ k^2\left[1-{1\over k^2\eta^2}\left(2-{m^2\over H^2}\right)\right]\chi_k = 0\;,
\end{eqnarray*}
where $k = \vert\vec k\rvert$.
The general solution of this equation can be expressed as a linear combination of Hankel functions: 
\begin{equation*}
\chi_k(\eta) = \sqrt{-k\eta}\left[A_k\,{\mathcal H}_\nu^{(1)}(-k\eta)+B_k\,{\mathcal H}_\nu^{(2)}(-k\eta)\right]\;,
\end{equation*}
where the index $\nu$ is 
\begin{equation}
\nu=\sqrt{{9\over4}-\frac{m^2}{H^2}}\;.
\label{index}
\end{equation}
The choice of the coefficients $A_k$ and $B_k$ defines a vacuum state $|0\rangle$ annihilated by $a_{\vec{k}}$:
\begin{eqnarray*}
a_{\vec{k}}|0\rangle=0\;\; \text{for any $\vec k$}\;.
\end{eqnarray*}
If one wants to have a vacuum that in the remote past $\eta\to-\infty$ (or, equivalently, for modes with very short physical wavelengths, $-kH\eta\gg H$) behaves like the vacuum in Minkowski spacetime, 
\begin{eqnarray*}
\chi_k(\eta)\to{e^{-ik\eta}\over\sqrt{2k}}\;,
\end{eqnarray*}
one should choose 
\begin{equation*}
\chi_k(\eta) = -{\sqrt\pi\over2}\sqrt{-\eta}\,{\mathcal H}_\nu^{(1)}\left(-k\eta\right)\;.
\end{equation*}
Such a choice is called the Bunch-Davies vacuum \cite{BD}. The mode functions of the original field, $\phi_k=a^{-1}\chi_k$, are then given by  
\begin{equation}
\phi_k(\eta) = -{\sqrt\pi\over2}H(-\eta)^{3/2}{\mathcal H}_\nu^{(1)}(-k\eta)\;.
\label{BDmodes}
\end{equation}
As long as $m>0$, this state is de Sitter invariant \cite{Allen:1985ux}.

Throughout this paper we will be assuming that $m^2\ll H^2$, in which case the order \eqref{index} of the Hankel function  becomes 
\begin{equation}
\nu\approx{3\over2}-u, \text{ with $u\equiv{m^2\over3H^2}\ll1$}\;.
\end{equation}

\section{Perturbative calculation of the two-point correlation function}
In this section we present a perturbative calculation of the long-wavelength part of the two-point function. There are two reasons why it is meaningful to consider exclusively the long-wavelength modes. 

The first reason is physical. The fluctuations relevant for the formation of the observed large-scale structure of the universe are those whose wavelength, by the end of inflation, has been stretched to a size much larger than the Hubble horizon. 

The second reason is mathematical. Calculations are much simpler if instead of the modes \eqref{BDmodes}, one uses their long-wavelength limit. At the same time, in many cases the results can reflect the behavior of the untruncated theory. In the small mass limit, the long-wavelength two-point function matches with the untruncated one for large separations and for coinciding spacetime points. For the free theory, this is explicitly shown at the end of Sec.~3.1. Perturbative corrections (which we calculate in Sections 3.3 and 3.4) consist of products of Wightman functions and retarded Green's functions, integrated over momenta and intermediate time coordinates. Here too the leading contribution in small mass derives entirely from the long-wavelength limit of these functions.                   

\subsection{The free theory}
At the level of the free theory, the two-point function in a vacuum state $|0\rangle$ defined by a set of modes $\phi_k(\eta)$  is given by
\begin{eqnarray}
\langle\phi(\vec x,t_1)\phi(\vec y,t_2)\rangle_{\lambda^0}&\equiv&\langle 0|\phi(\vec x,t_1)\phi(\vec y,t_2)|0\rangle\nonumber\\&=&\int\frac{d^3{\vec k}}{(2\pi)^3}\phi_k(\eta_1)\phi_k^*(\eta_2)e^{i\vec k\cdot(\vec x-\vec y)}\nonumber\\&=&{1\over2\pi^2}\int_0^\infty dk\;k^2{\sin{kr}\over kr}\phi_k(\eta_1)\phi_k^*(\eta_2)\,,\nonumber\\
\label{2point}
\end{eqnarray}
where on the last line the integration over the angles was performed, and $r\equiv\vert\vec x-\vec y\rvert$.

The long-wavelength part of \eqref{2point} consists of modes with physical momenta much less than $H$: 
\begin{eqnarray*}
{k\over a(\eta_1)H}=-k\eta_1<\epsilon\;,\;\; {k\over a(\eta_2)H}=-k\eta_2<\epsilon\;,
\end{eqnarray*}
where $\epsilon$ is a small constant parameter, $\epsilon\ll1$. Expanding the mode functions \eqref{BDmodes} in this limit yields
\begin{eqnarray}
\phi_k(\eta)&\approx&{iH\over\sqrt2}(-\eta)^{3/2}(-k\eta)^{-\nu}={iH\over\sqrt{2k^3}}(-k\eta)^{u}.
\label{longmodes}
\end{eqnarray}
The long-wavelength part of the two-point function is then given by 
\begin{eqnarray}
&&\langle\phi(\vec x,t_1)\phi(\vec y,t_2)\rangle_{\lambda^0,L}\nonumber\\&&~~~~~~~~={H^2(\eta_1\eta_2)^u\over4\pi^2}\int_{0}^{-\epsilon/\eta_{\rm m}}{dk\over k}{k^{2u}\sin(kr)\over kr}\;,
\label{2pointl}
\end{eqnarray}
where $\eta_{\rm m}$ is the earliest time that accompanies the momentum $\vec k$: $\eta_{\rm m}\equiv {\rm min}(\eta_1,\eta_2)$, and the subscript $L$ stands for ``long-wavelength''.

In the case of coinciding spacetime points, Eq.~\eqref{2pointl} gives
\begin{eqnarray}
\langle\phi^2(\vec x,t)\rangle_{\lambda^0,L}={H^2(-\eta)^{2u}\over4\pi^2}\int_{0}^{-\epsilon/\eta}{dk\over k^{1-2u}}={H^2\over8\pi^2}{\epsilon^{2u}\over u}.\nonumber\\
\label{2pointlc}
\end{eqnarray}
By expanding $\epsilon^{2u}$ for small values of $u$ 
\begin{equation*}
\epsilon^{2u}=1+2u\ln\epsilon+\mathcal{O}(u^2\ln^2\epsilon),
\label{epexp}
\end{equation*}
we see that if we choose $\epsilon$ in such a way that it satisfies the following conditions
\begin{equation}
\exp\left(-u^{-1}\right)\ll\epsilon\ll1,
\label{cond}
\end{equation}
then $\epsilon^{2u}$ in Eq.~\eqref{2pointlc} may be replaced by 1: 
\begin{eqnarray}
\langle\phi^2(\vec x,t)\rangle_{\lambda^0,L}={H^2\over8\pi^2 u}={3H^4\over8\pi^2m^2}\;.
\label{coinc}
\end{eqnarray}
It is very important that $\langle\phi^2\rangle_{\lambda^0,L}$ appears to be anomalously large---much larger than $H^2$. This makes $\langle\phi^2\rangle_{L}$ a very good approximation for the total renormalized value of
$\langle\phi^2\rangle_{\rm ren}$, both here and at higher loops, since small-scale and renormalization contributions to this quantity are of the order of $H^2$ (possibly up to a logarithmic multiplier).

To consider a more general case, it is convenient to introduce a new integration variable $w\equiv kr$ and rewrite Eq.~\eqref{2pointl} as  
\begin{eqnarray}
\langle\phi(\vec x,t_1)\phi(\vec y,t_2)\rangle_{\lambda^0,L}={H^2\over4\pi^2}\left({\eta_1\eta_2\over r^2}\right)^u\int_{0}^{-\epsilon r/\eta_{\rm m}}{\sin w\over w^{2-2u}}{dw}\;.\nonumber\\
\label{2pointln}
\end{eqnarray}

In order to analyze this expression, let us look at the following quantity: given two points in de Sitter spacetime, there is a de Sitter invariant function associated with them \cite{Allen:1985ux,Allen:1987tz}:
\begin{equation}
Z(X,Y)=-H^2\eta_{\mu\nu}X^{\mu}Y^{\nu},
\label{z}
\end{equation}
where $X$ and $Y$ represent coordinates in five-dimensional Minkowski embedding spacetime with the metric $\eta_{\mu\nu}={\rm diag}(1,-1,-1,-1,-1)$. We describe this function in more details in Appendix A. The explicit expression for \eqref{z} depends on the coordinate system. In spatially flat coordinates \eqref{conformal} that we use throughout this paper it has the form: 
\begin{eqnarray}
Z(\vec x_1,\eta_1;\vec y,\eta_2)={\eta_1^2+\eta_2^2-\vert\vec x-\vec y\rvert^2\over2\eta_1\eta_2}\;.
\label{z1}
\end{eqnarray}
The important property of this quantity is that it allows us to distinguish between timelike and spacelike related points. If points are timelike separated, then $Z>1$; if points are lightlike separated, then $Z=1$, and if points are spacelike separated, then $Z<1$. 

If $(\vec x,t_1)$ and $(\vec y,t_2)$ are related in such a way that
\begin{equation}
Z>1-{1\over2\epsilon^2}\;,
\label{Z1}
\end{equation}
then it can be deduced from
\begin{eqnarray*}
{\eta_1^2+\eta_2^2-\vert\vec x-\vec y\rvert^2\over2\eta_1\eta_2}>1-{1\over2\epsilon^2}\;
\end{eqnarray*}
that 
\begin{equation*}
{-r/\eta_{\rm m}}<{1/\epsilon}\;,
\end{equation*}
where $\eta_{\rm m}\equiv {\rm min}(\eta_1,\eta_2)$ and $r\equiv\vert\vec x-\vec y\rvert$. This means that the upper limit of the integral in \eqref{2pointln} is smaller than $1$, so we can use the approximation $\sin w\approx w$ and obtain
\begin{eqnarray}
\langle\phi(\vec x,t_1)\phi(\vec y,t_2)\rangle_{\lambda^0,L}&=&{H^2\over4\pi^2}\left({\eta_1\eta_2\over r^2}\right)^u\int_{0}^{-\epsilon r/\eta_{\rm m}}{dw\over w^{1-2u}}\nonumber\\&=&{H^2\over8\pi^2}{\epsilon^{2u}\over u}\bigg({\eta_1\eta_2\over\eta_{\rm m}^2}\bigg)^u\;.
\end{eqnarray}
As explained earlier, $\epsilon^{2u}$ can be replaced by $1$ if it satisfies the condition \eqref{cond}. Hence in terms of the cosmic time we have    
\begin{eqnarray}
\langle\phi(\vec x,t_1)\phi(\vec y,t_2)\rangle_{\lambda^0,L}&=&{H^2\over8\pi^2u}e^{-uH\vert t_1-t_2\rvert}\;.
\label{tl0}
\end{eqnarray}

There are two important subcases of \eqref{Z1} for which the two-point function is given by the above expression. One is when points $(\vec x,t_1)$ and $(\vec y,t_2)$ are timelike or lightlike separated; the other is when these points have coinciding time coordinates and the physical spatial distance between them satisfies $a(t)r<(\epsilon H)^{-1}$. In the latter instance, from Eq.~\eqref{tl0} we obtain the same result as in Eq.~\eqref{coinc}. This means that as far as the long-wavelength correlation function is concerned, there is no difference between coinciding spacetime points and points on a constant time hypersurface that are separated by a proper distance less than $(\epsilon H)^{-1}$.              

Turning to the case where the upper limit of the integral in \eqref{2pointln} is bigger than $1$, that is,
\begin{equation}
(-r/\eta_{\rm m})>1/\epsilon\;,
\label{slen}
\end{equation}
it can be shown by using \eqref{z1} that 
\begin{equation*}
Z<1-{1\over2\epsilon^2}\ll-1\;;
\end{equation*}
therefore, this corresponds to the regime of large spacelike separation between points. Let us split the integral in \eqref{2pointln} into two parts 
\begin{eqnarray} 
\int_{0}^{-\epsilon r/\eta_{\rm m}}{\sin w\over w^{2-2u}}{dw}=\!\!\int_{0}^{\infty}{\sin w\over w^{2-2u}}{dw}-\int_{-\epsilon r/\eta_{\rm m}}^{\infty}{\sin w\over w^{2-2u}}{dw}.\nonumber\\
\label{split}
\end{eqnarray}
The first integral can be calculated exactly
\begin{eqnarray*}
\int_{0}^{\infty}{\sin w\over w^{2-2u}}{dw}=-\cos(\pi u)\Gamma(2u-1)={1\over2u}\bigg(1+\mathcal{O}(u)\bigg)\;,\nonumber\\
\end{eqnarray*}
while the second integral can be shown to be smaller than $(1-2u)^{-1}$:  
\begin{eqnarray*}
\left\vert\int_{-\epsilon r/\eta_{\rm m}}^{\infty}{\sin w\over w^{2-2u}}{dw}\right\rvert<\int_{-\epsilon r/\eta_{\rm m}}^{\infty}{w^{2u-2}}{dw}<{1\over1-2u}\;,\nonumber\\
\end{eqnarray*}
where the second inequality follows from \eqref{slen}. Hence in the small mass limit ($u\ll1$), the second term on the right-hand side of \eqref{split} can be neglected, and we conclude that in the regime of large spacelike separation the two-point function equals      
\begin{eqnarray}
\langle\phi(\vec x,t_1)\phi(\vec y,t_2)\rangle_{\lambda^0,L}&=&{H^2\over8\pi^2u}\left({\eta_1\eta_2\over r^2}\right)^u\nonumber\\&=&{H^2\over8\pi^2u}e^{-uH(t_1+t_2)}(rH)^{-2u}\;.\nonumber\\
\label{tps}
\end{eqnarray}
As we can see, in this regime the equal-time two-point function,  
\begin{eqnarray}
\langle\phi(\vec x,t)\phi(\vec y,t)\rangle_{\lambda^0,L}={3H^4\over8\pi^2m^2}(RH)^{-{2m^2\over3H^2}}\;,
\end{eqnarray}
depends only on the physical spatial distance $R\equiv re^{Ht}$.

Let us compare the long-wavelength results we have obtained so far with the untruncated two-point function. This two-point function is calculated by using Eq.~\eqref{2point} with mode functions given by \eqref{BDmodes}. The result is well known \cite{BD,Allen:1987tz}: for spacelike separated points, it is equal to     
\begin{eqnarray}
&&\langle\phi(\vec x,t_1)\phi(\vec y,t_2)\rangle_{\lambda^0}\nonumber\\&&~~~~={H^2(1-u)(2-u)\over16\pi\sin\pi u}F\left(3-u,u;2;{1+Z\over2}\right),
\label{hyper}
\end{eqnarray}
where $F$ is a hypergeometric function, which is real for $Z<1$ and has a branch cut that runs from $Z=1$ to $+\infty$. To study \eqref{hyper} in the limit of coinciding spacetime points, we expand it around $Z=1$:
\begin{eqnarray}
\langle\phi(\vec x,t_1)\phi(\vec y,t_2)\rangle_{\lambda^0}\to{H^2\over8\pi^2}\bigg({1\over u}&-&2+{1\over1-Z}\nonumber\\&-&\ln{(1-Z)\over2}+\mathcal{O}(u)\bigg)\nonumber\\
\label{UV}
\end{eqnarray}
as $(\vec x,t_1)\to(\vec y,t_2)$; $\mathcal{O}(u)$ contains terms that are finite as well as terms that diverge as $Z\to1$. It is evident from the above expression that the term leading in $u$ is free of ultraviolet divergences and coincides with the result \eqref{coinc}, obtained by the long-wavelength truncation. 

When points are timelike separated, the untruncated correlation function equals 
\begin{eqnarray}
&&\langle\phi(\vec x,t_1)\phi(\vec y,t_2)\rangle_{\lambda^0}\nonumber\\&&~~~~={H^2(1-u)(2-u)\over16\pi\sin\pi u}F\left(3-u,u;2;{1+Z\pm i\epsilon\over2}\right),\nonumber\\
\label{hypert}
\end{eqnarray}
where $\epsilon$ is a positive real infinitesimal. The plus sign in front of $\epsilon$ refers to the case when $(\vec x,t_1)$ is in the past light cone of $(\vec y,t_2)$, and the minus sign to the case when $(\vec x,t_1)$ is in the forward light cone of $(\vec y,t_2)$. The $+i\epsilon$ and $-i\epsilon$ prescriptions push the branch cut in the complex $(1-Z)$-plane slightly above and slightly below the negative real axis, respectively. Consequently, the sign of the imaginary part of \eqref{hypert} depends on the sign of $t_1-t_2$: when $(t_1-t_2)>0$, the imaginary part of \eqref{hypert} is negative, and when $(t_1-t_2)<0$, the imaginary part of \eqref{hypert} is positive.              

It follows from Eqs.~\eqref{hyper} and \eqref{hypert} that for $|Z|\gg1$,   
\begin{eqnarray}
\langle\phi(\vec x,t_1)\phi(\vec y,t_2)\rangle_{\lambda^0}\to{H^2\over8\pi^2 u}\bigg(1+\mathcal{O}(u^0)\bigg)|Z|^{-u}\;.
\label{2pointZ}
\end{eqnarray}     
For large timelike separations, $Z$ is much greater than unity and can be approximated as $Z\approx 0.5{e^{H\vert t_1-t_2\rvert}}$. Therefore, \eqref{2pointZ} becomes
\begin{eqnarray*}
\langle\phi(\vec x,t_1)\phi(\vec y,t_2)\rangle_{\lambda^0}\to{H^2\over8\pi^2 u}e^{-uH\vert t_1-t_2\rvert}\;,
\end{eqnarray*}  
which matches with its long-wavelength counterpart \eqref{tl0}. For large spacelike separations, $Z\ll-1$ and, with the additional assumption that $-r/\eta_{\rm m}\gg1$, it can be approximated as $Z\approx-0.5H^2r^2e^{H(t_1+t_2)}$. Hence \eqref{2pointZ} reduces to
\begin{eqnarray*}
\langle\phi(\vec x,t_1)\phi(\vec y,t_2)\rangle_{\lambda^0}\to{H^2\over8\pi^2u}e^{-uH(t_1+t_2)}(rH)^{-2u}\;,
\end{eqnarray*}  
which also coincides with its long-wavelength analog \eqref{tps}.

Just as in flat spacetime, the expectation value of the commutator of two fields vanishes for spacelike separated points and is nonzero for timelike separated points. For example, if $(\vec x,t_1)$ is in the future of $(\vec y,t_2)$, then
\begin{eqnarray*}
&&\!\!\!\!\langle[\phi(\vec x,t_1),\phi(\vec y,t_2)]\rangle_{\lambda^0}\nonumber\\&&={iH^2(1-u)(2-u)\over8\pi\sin\pi u}{\rm Im}\left\{F\left(3-u,u;2;{1+Z-i\epsilon\over2}\right)\right\}.
\end{eqnarray*}
On the other hand, it follows from Eqs.~\eqref{tl0} and \eqref{tps} that       
\begin{eqnarray}
\langle[\phi(\vec x,t_1),\phi(\vec y,t_2)]\rangle_{\lambda^0,L}=0
\label{commlong}
\end{eqnarray}
both for timelike and spacelike related points. The vanishing of the commutator in \eqref{commlong} indicates that the long-wavelength part of the field in a sense behaves like a classical quantity. As we will see, the perturbative corrections don't change the value of the long-wavelength commutator: it remains equal to zero.

\subsection{The ``in-in'' formalism}
To calculate perturbative corrections to the two-point function, we will work in the interaction picture and use the ``in-in'' formalism \cite{Schwinger,Keldysh,Bakshi:1962dv,Jordan,Weinberg:2005vy}. In this formalism the two-point function can be written as
\begin{eqnarray}
&&\!\!\!\!\langle\phi(\vec x,t_1)\phi(\vec y,t_2)\rangle\nonumber\\&&\quad\quad\quad\quad=\bigl\langle 0\big|
U^\dagger(t_1,-\infty)\phi_I(\vec x,t_1)U(t_1,-\infty)\nonumber\\&&\quad\quad\quad\quad\quad~~~\times U^\dagger(t_2,-\infty)\phi_I(\vec y,t_2) 
U(t_2,-\infty)\big|0\bigr\rangle\;.\nonumber\\
\label{ctp}
\end{eqnarray}
Here, $|0\rangle$ is the vacuum state of the free theory and $U(t,-\infty)$ is the unitary time-evolution operator:
\begin{eqnarray*}
U(t,-\infty)&=&T e^{-i\int_{-\infty}^t dt'\, H_I(t')}\;,\\
U^\dagger(t,-\infty)&=&\overline{T} e^{i\int_{-\infty}^t dt'\, H_I(t')}\;,
\end{eqnarray*}
where $T$ stands for time-ordering, $\overline{T}$ --- for anti-time-ordering. Just as in ``in-out'' formalism, $\phi_I$-s are interaction picture fields with time evolution governed by the free theory Hamiltonian and $H_I$ is the interaction Hamiltonian in the interaction picture. In what follows we suppress the subscript $I$ of the interaction picture fields.

Unlike in ``in-out'' formalism, conditions on fields and states are not imposed at both very early and very late times, but only at very early times: both the vacuum state $|0\rangle$ and its Hermitian conjugate are defined at $-\infty$.

For practical calculations it will be convenient to rewrite Eq.~\eqref{ctp} in a slightly different form. Let us represent the time-evolution operator and its Hermitian conjugate as    
\begin{eqnarray*}
U(t_1,-\infty)&=&U^\dagger(\infty,t_1)U(\infty,-\infty)\;,\\
U^\dagger(t_2,-\infty)&=&U^\dagger(\infty,-\infty)U(\infty,t_2)\;.
\end{eqnarray*}
Using the right-hand side of these equalities to replace the product $U(t_1,-\infty)U^\dagger(t_2,-\infty)$ in \eqref{ctp} gives
\begin{eqnarray}
\langle\phi(\vec x,t_1)\phi(\vec y,t_2)\rangle&=&\bigl\langle 0\big|U^\dagger(t_1,-\infty)\phi(\vec x,t_1)U^\dagger(\infty,t_1)\nonumber\\&&~~~\times U(\infty,t_2)\phi(\vec y,t_2) 
U(t_2,-\infty)\big|0\bigr\rangle\nonumber\\&=&\bigl\langle 0\big| 
\overline{T}\bigl(\phi(\vec x,t_1) e^{i\int_{-\infty}^{\infty} dt'\, H_I(t')}\bigr)\nonumber\\&&~~~\times T\bigl(\phi(\vec y,t_2) e^{-i\int_{-\infty}^{\infty} dt'\, H_I(t')}\bigr)\big|0\bigr\rangle\,.\nonumber\\
\label{ctp1}
\end{eqnarray}
The second equality is obtained by noticing that on the right-hand side of the first equality the first three operators are anti-time-ordered and the last three operators are time-ordered. 

The right-hand side of \eqref{ctp1} can be interpreted in the following way: we start with the initial state at $-\infty$, evolve forward to $t_2$ where the operator $\phi(\vec y,t_2)$  is inserted, then continue to evolve forward to $\infty$; afterwards, we evolve backward to $t_1$ where the operator $\phi(\vec x,t_1)$ occurs, then continue to evolve back to $-\infty$. This is why the ``in-in'' formalism is also called ``closed-time-path'' formalism. This interpretation makes it possible to write \eqref{ctp1} in terms of a single time-ordered expression \cite{Jordan,Collins:2011mz}: label the fields on the forward-flowing part of the path with a ``$+$'' superscript, and the fields on the backward-flowing part of the path with a ``$-$'' superscript; thereby, \eqref{ctp1} can be written as
\begin{eqnarray}
&&\!\!\!\!\langle\phi(\vec x,t_1)\phi(\vec y,t_2)\rangle\nonumber\\&&=\bigl\langle0\big| 
T\bigl(\phi^-(\vec x,t_1)\phi^+(\vec y,t_2)e^{-i\int_{-\infty}^{\infty} dt'\;[H_I^+(t')-H_I^-(t')]}\bigr)\big|0\bigr\rangle\,,\nonumber\\
\label{ctp2} 
\end{eqnarray}
where $H_I^{\pm}(t)\equiv H_I[\phi^{\pm}(t,\vec{x})]$, and the time-ordering operation is extended in the following way: two ``$+$'' fields are ordered as usual,
\begin{eqnarray*}
T(\phi^+(\vec{x},t)\phi^+(\vec{y},t'))&=&\theta(t-t')\phi^+(\vec{x},t)\phi^+(\vec{y},t')\nonumber\\&&+\theta(t'-t)\phi^+(\vec{y},t')\phi^+(\vec{x},t)\;,
\end{eqnarray*}
``$-$'' fields always occur after ``$+$'' fields,
\begin{eqnarray*}
T(\phi^+(\vec{x},t)\phi^-(\vec{y},t'))&=&\phi^-(\vec{y},t')\phi^+(\vec{x},t)\;,\\
T(\phi^-(\vec{x},t)\phi^+(\vec{y},t'))&=&\phi^-(\vec{x},t)\phi^+(\vec{y},t')\;,
\end{eqnarray*}
and two ``$-$'' fields are ordered in the opposite of the usual sense,
\begin{eqnarray*}
T(\phi^-(\vec{x},t)\phi^-(\vec{y},t'))&=&\theta(t'-t)\phi^-(\vec{x},t)\phi^-(\vec{y},t')\nonumber\\&&+\theta(t-t')\phi^-(\vec{y},t')\phi^-(\vec{x},t)\;.\nonumber
\end{eqnarray*}

We can use Wick's theorem to express the time-ordered products in \eqref{ctp2} in terms of field contractions, but unlike in ``in-out'' formalism, there are four types of Wick contractions (and hence, four propagators)
\begin{eqnarray*}
&&\langle 0|T(\phi^+(\vec{x},t)\phi^+(\vec{y},t'))|0\rangle\\
&&~~=\theta(t-t')G^{>}(\vec x,t;\vec y,t')+\theta(t'-t)G^{<}(\vec x,t;\vec y,t')\;,\\
&&\langle 0|T(\phi^+(\vec{x},t')\phi^-(\vec{y},t'))|0\rangle=G^{<}(\vec x,t;\vec y,t')\;,\\
&&\langle 0|T(\phi^-(\vec{x},t)\phi^+(\vec{y},t'))|0\rangle=G^{>}(\vec x,t;\vec y,t')\;,\\
&&\langle 0|T(\phi^-(\vec{x},t)\phi^-(\vec{y},t'))|0\rangle\\
&&~~=\theta(t'-t)G^{>}(\vec x,t;\vec y,t')+\theta(t-t')G^{<}(\vec x,t;\vec y,t')\;,
\end{eqnarray*}
where $G^{>}(\vec x,t;\vec y,t')$ and $G^{<}(\vec x,t;\vec y,t')$ are the Wightman functions\\
\begin{eqnarray*}
G^{>}(\vec x,t;\vec y,t')&\equiv&\langle 0|\phi(\vec{x},t)\phi(\vec{y},t')|0\rangle\\
&=&\int\frac{d^3\vec{k}}{(2\pi)^3}\;e^{i{\vec{k}(\vec{x}-\vec{y})}}G_k^{>}(t,t')\;,\\
G^{<}(\vec x,t;\vec y,t')&\equiv&\langle 0|\phi(\vec{y},t')\phi(\vec{x},t)|0\rangle\\
&=&\int\frac{d^3\vec{k}}{(2\pi)^3}\;e^{i{\vec{k}(\vec{x}-\vec{y})}}G_k^{<}(t,t')\;.
\end{eqnarray*} 
In terms of the conformal time, the momentum representations of the Wightman functions are given by
\begin{eqnarray}
&&\!\!\!\!G_k^>(\eta,\eta')=\phi_k(\eta)\phi_k^*(\eta')\nonumber\\&&={\pi H^2\over4}(-\eta)^{3/2}(-\eta')^{3/2}{\mathcal H}_\nu^{(1)}(-k\eta){\mathcal H}_\nu^{(2)}(-k\eta')\;,
\label{Wight1}
\end{eqnarray}
\begin{eqnarray}
G_k^<(\eta,\eta')=\phi_k^*(\eta)\phi_k(\eta')=G_k^>(\eta',\eta)\;, 
\label{Wight2}
\end{eqnarray}
where we have used the mode functions \eqref{BDmodes} associated with the Bunch-Davies state.

\subsection{The one-loop correction}
The interaction Hamiltonian of the scalar field with quartic self-interaction is given by
\begin{eqnarray*}
H_I(t)=\int d^3\vec{x}\;\left(a^3(t){\lambda\over4}\phi^4\right)\;.
\end{eqnarray*}
The one-loop correction to the two-point function can be obtained by expanding Eq.~\eqref{ctp2} to first order in the coupling constant $\lambda$:  
\begin{eqnarray*}  
&&\langle\phi(\vec x,t_1)\phi(\vec y,t_2)\rangle_{\lambda}\nonumber\\&&~~=-i\int_{-\infty}^{\infty} dt'\;\langle0|T\bigg(\phi^-(\vec x,t_1)\phi^+(\vec y,t_2)\nonumber\\&&\quad\quad\quad\quad\quad\quad\quad\quad\quad~\times\Big[H_I^+(t')-H_I^-(t')\Big]\bigg)|0\rangle\;.
\end{eqnarray*}
\setlength{\unitlength}{1mm}
\thicklines
\begin{picture}(90,30)
\put(18,15){\circle*{2}}
\put(14,11){$(\vec{x},t_1)$}
\put(18,15){\line(1,0){22}}
\put(40,15){\circle*{2}}
\put(40,21.5){\circle{13}}
\put(39,11){$t'$}
\put(40,15){\line(1,0){22}}
\put(62,15){\circle*{2}}
\put(58,11){$(\vec{y},t_2)$}
\put(36,5){Fig.1}
\end{picture}
Taking the contractions that lead to the connected diagram (Fig.1) and switching from the cosmic time $t$ to the conformal time $\eta$  yields  
\begin{eqnarray}
&&\!\!\!\!\!\!\!\!\langle\phi(\vec x,t_1)\phi(\vec y,t_2)\rangle_{\lambda}\nonumber\\&=&-{3i\lambda}\int_{-\infty}^{0}\!\!{d\eta'}a^4(\eta')\int\!\!\frac{d^3{\vec k}}{(2\pi)^3}{e^{i\vec k\cdot(\vec x-\vec y)}}\nonumber\\&&\times\bigg[\theta(\eta_1-\eta')G_k^C(\eta_1,\eta')G_k^<(\eta_2,\eta')\nonumber\\&&+\theta(\eta_2-\eta')G_k^C(\eta_2,\eta')G_k^>(\eta_1,\eta')\bigg]\!\!\int\!\!\frac{d^3{\vec \ell}}{(2\pi)^3}G_\ell^>(\eta',\eta'),\nonumber\\
\label{1l}
\end{eqnarray}
where $G_k^C(\eta,\eta')\equiv G_k^>(\eta,\eta')-G_k^<(\eta,\eta')$. Note that the product $\theta(\eta-\eta')G_k^C(\eta,\eta')$ is nothing but the momentum representation of the retarded Green's function. 

To evaluate the long-wavelength part of \eqref{1l}, let us look at the small (physical) momenta expansion of the Wightman functions. It follows from Eq.~\eqref{longmodes} that    
\begin{equation}
G_k^>(\eta,\eta')\approx{H^2\over2k^3}(-k\eta)^u(-k\eta')^u\approx G_k^<(\eta,\eta')\;.
\label{Wl}
\end{equation}
The long-wavelength part of the integral over $\vec\ell$ in Eq.~\eqref{1l} is equal to $\langle\phi^2\rangle_{\lambda^0,L}$ (cf. Eq~\eqref{2pointlc}):
\begin{eqnarray}
\int^{-\epsilon/\eta'}\!\!\frac{d^3{\vec \ell}}{(2\pi)^3}G_\ell^>(\eta',\eta')&\approx&{H^2\over4\pi^2}\int_{0}^{-\epsilon/\eta'}\!\!{d\ell\over\ell}(-\ell\eta')^{2u}\nonumber\\&=&{H^2\over8\pi^2}{\epsilon^{2u}\over u}\;.
\label{WI}
\end{eqnarray}   
In order to extract the leading terms from the difference of the Wightman functions, it is convenient to rewrite it in terms of the Bessel functions of the first and second kind; from Eqs.~\eqref{Wight1} and \eqref{Wight2} we have   
\begin{eqnarray*}
G_k^C(\eta,\eta')&=&i{\pi H^2\over2}(-\eta)^{3/2}(-\eta')^{3/2}\nonumber\\&&\times\biggl[Y_{\nu}(-k\eta)J_{\nu}(-k\eta')-J_{\nu}(-k\eta)Y_{\nu}(-k\eta')\biggr],
\end{eqnarray*}
which in the long-wavelength limit gives
\begin{eqnarray*}
G_k^C(\eta,\eta')&\approx&-{i\over2\nu}H^2(-\eta)^{3/2}(-\eta')^{3/2}\nonumber\\&&\times\biggl[(-k\eta)^{-\nu}(-k\eta')^{\nu}-(-k\eta)^{\nu}(-k\eta')^{-\nu}\biggr]\\&\approx&-{i\over3}H^2(-\eta)^{u}(-\eta')^{3-u}+{i\over3}H^2(-\eta)^{3-u}(-\eta')^{u}.
\end{eqnarray*}
In what follows we will only keep the first term of this approximation,
\begin{eqnarray}
G_k^C(\eta,\eta')&\approx&-{i\over3}H^2(-\eta)^{u}(-\eta')^{3-u}\;,
\label{WCl}
\end{eqnarray}
since (as it will become clear from the calculations below) the part of the two-point function that would arise due to the second term would be suppressed by an extra factor of $u$. 

Substituting Eqs.~\eqref{Wl}, \eqref{WI} and \eqref{WCl} into Eq.~\eqref{1l} gives         
\begin{eqnarray}
&&\!\!\!\!\!\!\!\!\langle\phi(\vec x,t_1)\phi(\vec y,t_2)\rangle_{\lambda,L}\nonumber\\&=&{\lambda H^2\over16\pi^2}{\epsilon^{2u}\over u}(\eta_1\eta_2)^{u}\bigg\{\int_{-\infty}^{\eta_1}{d\eta'\over\eta'}\int^{-\epsilon/\eta_{\rm m}}\frac{d^3{\vec k}}{(2\pi)^3}{e^{i\vec k\cdot(\vec x-\vec y)}\over k^{3-2u}}\nonumber\\
&&\quad\quad\quad\quad\quad\quad\quad+\int_{-\infty}^{\eta_2}{d\eta'\over\eta'}\int^{-\epsilon/\eta_{\rm m}}\frac{d^3{\vec k}}{(2\pi)^3}{e^{i\vec k\cdot(\vec x-\vec y)}\over k^{3-2u}}\bigg\}\nonumber\\
&=&{\lambda H^2\over32\pi^4}{\epsilon^{2u}\over u}\left({\eta_1\eta_2\over r^2}\right)^u\!\bigg\{\int_{-\infty}^{\eta_1}{d\eta'\over\eta'}\int_0^{-\epsilon r/\eta_{\rm m}}{\sin w\over w^{2-2u}}{dw}\nonumber \\
&&\quad\quad\quad\quad\quad\quad\quad\quad+\int_{-\infty}^{\eta_2}{d\eta'\over\eta'}\int_0^{-\epsilon r/\eta_{\rm m}}{\sin w\over w^{2-2u}}{dw}\bigg\}\;,\nonumber\\
\label{1ll}
\end{eqnarray}
where $\eta_{\rm m}$ is the earliest time that accompanies the momentum $\vec k$: $\eta_{\rm m}\equiv {\rm min}(\eta_1,\eta_2,\eta')$. This momentum cutoff is chosen in such a way as to ensure that all the physical scales associated with the comoving scale $k$ are smaller than $\epsilon H$ (that is, $-k\eta_1<\epsilon$, $-k\eta_2<\epsilon$ and $-k\eta'<\epsilon$). To make the calculations more tractable, let us assume that $t_1\geq t_2$ (that is, $\eta_1\geq\eta_2$), in which case $\eta_{\rm m}={\rm min}(\eta_2,\eta')$. In the second term in curly brackets, $\eta_{\rm m}$ is then equal to $\eta'$. In the first term, $\eta_{\rm m}$ is determined by splitting the time integral into two parts      
\begin{eqnarray}
\int_{-\infty}^{\eta_1}{d\eta'\over\eta'}\int_0^{-\epsilon r/\eta_{\rm m}}{\sin w\over w^{2-2u}}{dw}&&\nonumber\\&&\!\!\!\!\!\!\!\!\!\!\!\!\!\!\!\!\!\!\!\!\!\!\!\!\!\!\!\!\!\!\!\!\!\!\!\!=\int_{-\infty}^{\eta_2}{d\eta'\over\eta'}\!\int_0^{-\epsilon r/\eta'}\!{\sin w\over w^{2-2u}}{dw}\nonumber \\
&&\!\!\!\!\!\!\!\!\!\!\!\!\!\!\!\!\!\!\!\!\!\!\!\!\!\!\!\!+\int_{\eta_2}^{\eta_1}{d\eta'\over\eta'}\!\int_0^{-\epsilon r/\eta_2}\!{\sin w\over w^{2-2u}}{dw}\;;  
\end{eqnarray}
hence, Eq.~\eqref{1ll} becomes
\begin{eqnarray}
&&\!\!\!\!\!\!\!\!\langle\phi(\vec x,t_1)\phi(\vec y,t_2)\rangle_{\lambda,L}\nonumber\\&=&{\lambda H^2\over32\pi^4}{\epsilon^{2u}\over u}
\left({\eta_1\eta_2\over r^2}\right)^u\bigg\{2\int_{-\infty}^{\eta_2}{d\eta'\over\eta'}\!\int_0^{-\epsilon r/\eta'}\!{\sin w\over w^{2-2u}}{dw}\nonumber \\
&&\quad\quad\quad\quad\quad\quad\quad\quad~+\int_{\eta_2}^{\eta_1}{d\eta'\over\eta'}\!\int_0^{-\epsilon r/\eta_2}\!{\sin w\over w^{2-2u}}{dw}\bigg\}\;.\nonumber\\
\label{1loop}
\end{eqnarray} 

If the points $(\vec x,t_1)$ and $(\vec y,t_2)$ are related in such a way that $Z>1-(2\epsilon^{2})^{-1}$, the upper limit of the integral over $w$ in the second term in curly brackets is less than unity. Since $\eta'$ in the first term is less than $\eta_2$, the upper limit of the integral over $w$ in here is also less than unity. Hence in both terms $\sin{w}$ can be replaced by $w$, 
\begin{equation*} 
\int_{0}^{-\epsilon r/\eta}{\sin w\over w^{2-2u}}{dw}\approx\int_{0}^{-\epsilon r/\eta}{dw\over w^{1-2u}}={\epsilon^{2u}\over2u}\left({r\over-\eta}\right)^{2u}\;,
\end{equation*}
after which we obtain       
\begin{eqnarray*}
&&\langle\phi(\vec x,t_1)\phi(\vec y,t_2)\rangle_{\lambda,L}\nonumber\\&&~~~~=-{\lambda H^2\over64\pi^4}{\epsilon^{4u}\over u^3}\left({\eta_1\over\eta_2}\right)^{u}\left(1-u\ln{\eta_1\over\eta_2}\right)\nonumber\\&&~~~~=-{\lambda H^2\over64\pi^4}{\epsilon^{4u}\over u^3}\bigg(1+uH(t_1-t_2)\bigg)e^{-uH(t_1-t_2)}\;.
\end{eqnarray*}
Had we assumed that $t_2>t_1$, this result would still be true but with $t_1$ and $t_2$ exchanged. Therefore it can be generalized as      
\begin{eqnarray}
&&\langle\phi(\vec x,t_1)\phi(\vec y,t_2)\rangle_{\lambda,L}\nonumber\\&&~~~~=-{\lambda H^2\over64\pi^4u^3}\bigg(1+uH\vert t_1-t_2\rvert\bigg)e^{-uH\vert t_1-t_2\rvert}\;.\nonumber\\
\label{1loopt}
\end{eqnarray}
Specifically, in the case of coinciding spacetime points one finds 
\begin{equation}
\langle\phi^2(\vec x,t)\rangle_{\lambda,L}=-{27\lambda H^8\over64\pi^4m^6}\;.
\label{coinc-oneloop}
\end{equation}
Note that even though the second term in the parenthesis of \eqref{1loopt} has an additional factor of $u$, this term cannot be neglected because it is greater than unity when $\vert t_1-t_2\rvert>1/u$.       

As explained in Sec.~3.1, when $-\epsilon r/\eta_2>1$ (and consequently $Z<1-{1\over2\epsilon^2}$), the integral over $w$ in the second term in the curly brackets of Eq~\eqref{1loop} can be approximated as
\begin{eqnarray} 
\int_{0}^{-\epsilon r/\eta_2}{\sin w\over w^{2-2u}}{dw}\approx\int_{0}^{\infty}{\sin w\over w^{2-2u}}{dw}={1\over2u}\bigg(1+\mathcal{O}(u)\bigg)\;.\nonumber\\
\label{wint}
\end{eqnarray}
Performing the $\eta'$ integration then produces
\begin{eqnarray}
\int_{\eta_2}^{\eta_1}{d\eta'\over\eta'}\!\int_0^{-\epsilon r/\eta_2}\!{\sin w\over w^{2-2u}}{dw}\approx{1\over2u}\ln\left({\eta_1\over\eta_2}\right)\;.
\label{tint}
\end{eqnarray} 
To evaluate the first term, we split the time integral as follows:
\begin{eqnarray*}
2\int_{-\infty}^{\eta_2}{d\eta'\over\eta'}\!\int_0^{-\epsilon r/\eta'}\!{\sin w\over w^{2-2u}}{dw}&&\nonumber\\&&\!\!\!\!\!\!\!\!\!\!\!\!\!\!\!\!\!\!\!\!\!\!\!\!\!\!\!\!\!\!\!\!\!\!\!\!=2\int_{-\infty}^{-\epsilon r}{d\eta'\over\eta'}\!\int_0^{-\epsilon r/\eta'}\!{\sin w\over w^{2-2u}}{dw}\nonumber\\&&\!\!\!\!\!\!\!\!\!\!\!\!\!\!\!\!\!\!\!\!\!\!\!\!\!\!\!\!\!\!\!+2\int_{-\epsilon r}^{\eta_2}{d\eta'\over\eta'}\!\int_0^{-\epsilon r/\eta'}\!{\sin w\over w^{2-2u}}{dw}\;.\nonumber\\
\end{eqnarray*}
The second term on the right side of the above equality should be treated the same way as integrals \eqref{wint} and \eqref{tint},
\begin{eqnarray}
2\int_{-\epsilon r}^{\eta_2}{d\eta'\over\eta'}\!\int_0^{-\epsilon r/\eta'}\!{\sin w\over w^{2-2u}}{dw}\approx{1\over u}\ln{\left(-\eta_2\over\epsilon r\right)}\;,
\label{largew}
\end{eqnarray}
since the ratio $-\epsilon r/\eta'$ in here stays greater than unity. In the first term this ration stays smaller than unity, and hence this term can be calculated as 
\begin{eqnarray}
&&2\int_{-\infty}^{-\epsilon r}{d\eta'\over\eta'}\!\int_0^{-\epsilon r/\eta'}\!{\sin w\over w^{2-2u}}{dw}\nonumber\\
&&\quad\quad\quad\approx2\int_{-\infty}^{-\epsilon r}{d\eta'\over\eta'}\!\int_0^{-\epsilon r/\eta'}\!{dw\over w^{1-2u}}=-{1\over2u^2}\;.
\label{smallw}
\end{eqnarray}
Upon substituting Eqs.~\eqref{tint}, \eqref{largew} and \eqref{smallw} into Eq.~\eqref{1loop}, we obtain the long-wavelength correlation function at large spacelike separations:  
\begin{eqnarray*}
&&\langle\phi(\vec x,t_1)\phi(\vec y,t_2)\rangle_{\lambda,L}\nonumber\\&&\quad\quad=-{\lambda H^2\over64\pi^4}{\epsilon^{2u}\over u^3}\left({\eta_1\eta_2\over r^2}\right)^u\left[1+u\ln\left({\epsilon^2r^2\over\eta_1\eta_2}\right)\right]\;,
\end{eqnarray*}
which in terms of cosmic time becomes
\begin{eqnarray}
\langle\phi(\vec x,t_1)\phi(\vec y,t_2)\rangle_{\lambda,L}&=&-{\lambda H^2\over64\pi^4u^3}\left(r^2H^2e^{H(t_1+t_2)}\right)^{-u}\nonumber \\
&&\times\bigg\{1+u\ln\left({r^2H^2e^{H(t_1+t_2)}}\right)\bigg\}\;.\nonumber\\
\label{1loops}
\end{eqnarray}
If we considered the case $t_2>t_1$, this result would still be true.

The comparison of Eqs.~\eqref{1loopt} and \eqref{1loops} with \eqref{tl0} and \eqref{tps} reveals that the effective parameter of the perturbative expansion is not $\lambda$ but $\lambda/u^2$, so the perturbation theory is valid as long as $\lambda\ll m^4/H^4$. 
 
\subsection{The two-loop correction}
Upon expanding \eqref{ctp2} to second order in $\lambda$ and taking all of the possible field contractions that produce connected graphs, we obtain several diagrams with different topologies that contribute to the two-point function at $\lambda^2$-order.
\subsubsection{Two independent loops}
\setlength{\unitlength}{1mm}
\thicklines
\begin{picture}(250,30)
\put(4,15){\circle*{2}}
\put(0,11){$(\vec{x},t_1)$}
\put(4,15){\line(1,0){20}}
\put(24,15){\circle*{2}}
\put(24,20){\circle{10}}
\put(24,11){$t'$}
\put(24,15){\line(1,0){20}}
\put(44,15){\circle*{2}}
\put(44,20){\circle{10}}
\put(44,11){$t''$}
\put(44,15){\line(1,0){20}}
\put(64,15){\circle*{2}}
\put(60,11){$(\vec{y},t_2)$}
\put(32,5){Fig.2}
\end{picture} 
Taking the field contractions that correspond to the diagram with two independent loops (Fig.2), one finds
\begin{eqnarray*}
&&\!\!\!\!\!\!\!\!\langle\phi(\vec x,t_1)\phi(\vec y,t_2)\rangle_{\lambda^2}^{(1)}\\
&=&-9\lambda^2\bigg\{\int_{-\infty}^{0}\!\!d\eta'a^4(\eta')\int_{-\infty}^{\eta'}\!\!d\eta''a^4(\eta'')\\
&&\times\int\frac{d^3{\vec k}}{(2\pi)^3}{e^{i\vec k\cdot(\vec x-\vec y)}}\bigg[\theta(\eta_1-\eta')G_k^C(\eta_1,\eta')G_k^<(\eta_2,\eta'')\nonumber\\&&~~~~+\theta(\eta_2-\eta')G_k^C(\eta_2,\eta')G_k^>(\eta_1,\eta'')\bigg]G_k^C(\eta',\eta'')\nonumber\\&&+\int_{-\infty}^{\eta_1}\!\!{d\eta'}a^4(\eta')\int_{-\infty}^{\eta_2}\!\!{d\eta''}a^4(\eta'')\int\frac{d^3{\vec k}}{(2\pi)^3}{e^{i\vec k\cdot(\vec x-\vec y)}}\nonumber \\
&&\times G_k^C(\eta_1,\eta')G_k^C(\eta_2,\eta'')G_k^>(\eta',\eta'')\bigg\}\nonumber\\&&\times\!\int\!\frac{d^3{\vec\ell}}{(2\pi)^3}G_\ell^>(\eta',\eta')\!\int\!\frac{d^3{\vec p}}{(2\pi)^3}G_p^>(\eta'',\eta'').
\end{eqnarray*}
Using the small momenta approximations \eqref{Wl}, \eqref{WI} and \eqref{WCl} gives 
\begin{eqnarray}
&&\!\!\!\!\langle\phi(\vec x,t_1)\phi(\vec y,t_2)\rangle_{\lambda^2,L}^{(1)}={\lambda^2H^2\over256\pi^6}{\epsilon^{4u}\over u^2}\left({\eta_1\eta_2\over r^2}\right)^{u}\nonumber \\
&&~~\times\bigg\{\int_{-\infty}^{\eta_1}{d\eta'\over\eta'}\int_{-\infty}^{\eta'}{d\eta''\over\eta''}+\int_{-\infty}^{\eta_2}{d\eta'\over\eta'}\int_{-\infty}^{\eta'}{d\eta''\over\eta''}\nonumber\\&&~~\quad\quad+\int_{-\infty}^{\eta_1}{d\eta'\over\eta'}\int_{-\infty}^{\eta_2}{d\eta''\over\eta''}\bigg\}\int_0^{-\epsilon r/\eta_{\rm m}}{\sin w\over w^{2-2u}}{dw}\,,
\label{2loop}
\end{eqnarray}
where $\eta_{\rm m}$ is the earliest time that accompanies the momentum $\vec k$: $\eta_{\rm m}\equiv {\rm min}(\eta_1,\eta_2,\eta',\eta'')$. We can evaluate \eqref{2loop} by going through steps similar to the ones in the previous subsection; it is slightly more tedious here since splitting the time integrals produces more terms.    

When $Z>1-(2\epsilon^{2})^{-1}$, we obtain
\begin{eqnarray}
&&\!\!\!\!\langle\phi(\vec x,t_1)\phi(\vec y,t_2)\rangle_{\lambda^2,L}^{(1)}={\lambda^2H^2\over512\pi^6u^5}e^{-uH\vert t_1-t_2\rvert}\nonumber\\&&\quad\quad\times\bigg(1+uH\vert t_1-t_2\rvert+{1\over2}u^2H^2\vert t_1-t_2\rvert^2\bigg)\,,
\end{eqnarray}
which for coinciding spacetime points becomes 
\begin{eqnarray}
\langle\phi^2(\vec x,t)\rangle_{\lambda^2,L}^{(1)}={243\lambda^2H^{12}\over512\pi^6m^{10}}\;.
\label{2lc}
\end{eqnarray}
In $Z<1-(2\epsilon^{2})^{-1}$ regime, with the additional assumption that $-\epsilon r/{\rm min}(\eta_1,\eta_2)>1$, evaluating \eqref{2loop} yields
\begin{eqnarray}
&&\langle\phi(\vec x,t_1)\phi(\vec y,t_2)\rangle_{\lambda^2,L}^{(1)}={\lambda^2H^2\over512\pi^6u^5}\left(r^2H^2e^{H(t_1+t_2)}\right)^{-u}
\nonumber\\&&\quad\quad\quad\quad\times\bigg\{1+u\ln\left({r^2H^2e^{H(t_1+t_2)}}\right)\nonumber\\&&\quad\quad\quad\quad\quad~+{1\over2}u^2\ln^2\left({r^2H^2e^{H(t_1+t_2)}}\right)\bigg\}.
\end{eqnarray}
\subsubsection{Snowman diagram}
\setlength{\unitlength}{1mm}
\thicklines
\begin{picture}(90,40)
\put(10,15){\circle*{2}}
\put(6,11){$(\vec{x},t_1)$}
\put(10,15){\line(1,0){15}}
\put(25,15){\circle*{2}}
\put(24,11){$t'$}
\put(25,20){\circle{10}}
\put(25,25){\circle*{2}}
\put(24,21){$t''$}
\put(25,30){\circle{10}}
\put(25,15){\line(1,0){15}}
\put(40,15){\circle*{2}}
\put(36,11){$(\vec{y},t_2)$}
\put(21,5){Fig. 3}
\end{picture}
Next, we consider the snowman diagram (Fig.3). Contractions corresponding to this diagram produce the following result:
\begin{eqnarray*}
&&\!\!\!\!\!\!\!\!\langle\phi(\vec x,t_1)\phi(\vec y,t_2)\rangle_{\lambda^2}^{(2)}\nonumber\\&=&-9\lambda^2\int_{-\infty}^{0}\!\!{d\eta'}a^4(\eta')\int\frac{d^3{\vec k}}{(2\pi)^3}{e^{i\vec k\cdot(\vec x-\vec y)}}\nonumber \\
&&\times\bigg[\theta(\eta_1-\eta')G_k^C(\eta_1,\eta')G_k^<(\eta_2,\eta')\\
&&~~+\theta(\eta_2-\eta')G_k^C(\eta_2,\eta')G_k^>(\eta_1,\eta')\bigg]\int_{-\infty}^{\eta'}{d\eta''}a^4(\eta'')\nonumber\\&&\times\int\frac{d^3{\vec p}}{(2\pi)^3}\bigg(G^>_p(\eta',\eta'')+G^<_p(\eta',\eta'')\bigg)G_p^C(\eta',\eta'')\\
&&\times\int\frac{d^3{\vec\ell}}{(2\pi)^3}G_\ell^>(\eta'',\eta'')\;.\nonumber\\
\end{eqnarray*}
The long-wavelength approximation of this expression has the form:
\begin{eqnarray}
&&\!\!\!\!\langle\phi(\vec x,t_1)\phi(\vec y,t_2)\rangle_{\lambda^2,L}^{(2)}={\lambda^2H^2\over64\pi^6}{\epsilon^{2u}\over u}\left(\eta_1\eta_2\over r^2\right)^{u}\nonumber\\&&~~\times\bigg[\int_{-\infty}^{\eta_1}{d\eta'\over\eta'}(-\eta')^{2u}+\int_{-\infty}^{\eta_2}{d\eta'\over\eta'}(-\eta')^{2u}\bigg]\nonumber\\&&~~\times\int_0^{-\epsilon r/\eta_{\rm m}}{\sin w\over w^{2-2u}}{dw}\int_{-\infty}^{\eta'}{d\eta''\over\eta''}\int_0^{-\epsilon/\eta''}{dp\over p^{1-2u}}\;,
\label{2loopS}
\end{eqnarray}
where, as before, $\eta_{\rm m}$ is the earliest time that accompanies the momentum $\vec k$: $\eta_{\rm m}\equiv {\rm min}(\eta_1,\eta_2,\eta')$. Evaluating this in the case $Z>1-(2\epsilon^{2})^{-1}$ gives   
\begin{eqnarray}
&&\langle\phi(\vec x,t_1)\phi(\vec y,t_2)\rangle_{\lambda^2,L}^{(2)}\nonumber \\
&&~~~~={\lambda^2H^2\over512\pi^6u^5}\bigg(1+uH\vert t_1-t_2\rvert\bigg)e^{-uH\vert t_1-t_2\rvert}\;,
\end{eqnarray}
which for coinciding spacetime points reduces to 
\begin{eqnarray}
\langle\phi^2(\vec x,t)\rangle_{\lambda^2,L}^{(2)}={243\lambda^2H^{12}\over512\pi^6m^{10}}\;.
\label{smc}
\end{eqnarray}
On the other hand, when $Z<1-(2\epsilon^{2})^{-1}$, Eq.~\eqref{2loopS} gives  
\begin{eqnarray}
\langle\phi(\vec x,t_1)\phi(\vec y,t_2)\rangle_{\lambda^2,L}^{(2)}&=&{\lambda^2H^2\over512\pi^6u^5}\left(r^2H^2e^{H(t_1+t_2)}\right)^{-u}\nonumber\\&&\times\left\{1+u\ln\left({r^2H^2e^{H(t_1+t_2)}}\right)\right\}\;.\nonumber\\
\end{eqnarray}
\subsubsection{Sunset diagram}
\setlength{\unitlength}{1mm}
\thicklines
\begin{picture}(180,40)
\put(10,30){\circle*{2}}
\put(7,25){$(\vec{x},t_1)$}
\put(10,30){\line(1,0){20}}
\put(30,30){\circle*{2}}
\put(27,25){$t'$}
\put(30,30){\line(1,0){14}}
\put(44,30){\circle*{2}}
\put(44,25){$t''$}
\put(44,30){\line(1,0){20}}
\put(64,30){\circle*{2}}
\put(61,25){$(\vec{y},t_2)$}
\put(37,30){\circle{50}}
\put(32,15){Fig. 4}
\end{picture}
For the field contractions that correspond to the sunset diagram (Fig.4) we obtain the following expression:
\begin{eqnarray}
&&\!\!\!\!\!\!\!\!\langle\phi(\vec x,t_1)\phi(\vec y,t_2)\rangle_{\lambda^2}^{(3)}\nonumber \\
&=&-6\lambda^2\bigg\{\int_{-\infty}^{0}{d\eta'}a^4(\eta')\int_{-\infty}^{\eta'}{d\eta''}a^4(\eta'')\nonumber \\
&&\times\int\frac{d^3{\vec k}}{(2\pi)^3}\frac{d^3{\vec q}}{(2\pi)^3}\frac{d^3{\vec p}}{(2\pi)^3}\frac{d^3{\vec \ell}}{(2\pi)^3}\nonumber\\&&\times(2\pi)^3\delta^3(\vec k+\vec p+\vec l+\vec q){e^{i\vec k\cdot(\vec x-\vec y)}}
\nonumber\\
&&\times\bigg[\theta(\eta_1-\eta')G_k^C(\eta_1,\eta')G_k^<(\eta_2,\eta'')\nonumber \\
&&~~~+\theta(\eta_2-\eta')G_k^C(\eta_2,\eta')G_k^>(\eta_1,\eta'')\bigg]\nonumber\\
&&\times\bigg[G_p^>(\eta',\eta'')G_q^>(\eta',\eta'')+G_p^>(\eta',\eta'')G_q^<(\eta',\eta'')\nonumber \\
&&~~~+G_p^<(\eta',\eta'')G_q^<(\eta',\eta'')\bigg]G_\ell^C(\eta',\eta'')\!\!\!\!\!\!\nonumber\\&&+\int_{-\infty}^{\eta_1}{d\eta'}a^4(\eta')\int_{-\infty}^{\eta_2}{d\eta''}a^4(\eta'')\nonumber \\
&&\times\int\frac{d^3{\vec k}}{(2\pi)^3}\frac{d^3{\vec q}}{(2\pi)^3}\frac{d^3{\vec p}}{(2\pi)^3}\frac{d^3{\vec \ell}}{(2\pi)^3}\nonumber\\&&\times(2\pi)^3\delta^3(\vec k+\vec p+\vec l+\vec q){e^{i\vec k\cdot(\vec x-\vec y)}}G_k^C(\eta_1,\eta')\nonumber\\&&\times G_k^C(\eta_2,\eta'')G_\ell^>(\eta',\eta'')G_p^>(\eta',\eta'')G_q^>(\eta',\eta'')\bigg\}\;.
\label{sunset}
\end{eqnarray}
Let us use the delta function on the fourth line to perform the integral over $\vec\ell$ and the delta function on the second to last line to perform the integral over $\vec k$:
\begin{eqnarray*}
&&\!\!\!\!\!\!\!\!\langle\phi(\vec x,t_1)\phi(\vec y,t_2)\rangle_{\lambda^2}^{(3)}\nonumber \\
&=&-6\lambda^2\bigg\{\int_{-\infty}^{0}{d\eta'}a^4(\eta')\int_{-\infty}^{\eta'}{d\eta''}a^4(\eta'')\nonumber \\
&&\times\int\frac{d^3{\vec k}}{(2\pi)^3}\frac{d^3{\vec q}}{(2\pi)^3}\frac{d^3{\vec p}}{(2\pi)^3}{e^{i\vec k\cdot(\vec x-\vec y)}}\nonumber\\
&&\times\bigg[\theta(\eta_1-\eta')G_k^C(\eta_1,\eta')G_k^<(\eta_2,\eta'')\\
&&~~~+\theta(\eta_2-\eta')G_k^C(\eta_2,\eta')G_k^>(\eta_1,\eta'')\bigg]\nonumber\\
&&\times\bigg[G_p^>(\eta',\eta'')G_q^>(\eta',\eta'')+G_p^>(\eta',\eta'')G_q^<(\eta',\eta'')\\
&&~~~+G_p^<(\eta',\eta'')G_q^<(\eta',\eta'')\bigg]G_\ell^C(\eta',\eta'')\!\!\!\!\!\!\nonumber\\&&+\int_{-\infty}^{\eta_1}{d\eta'}a^4(\eta')\int_{-\infty}^{\eta_2}{d\eta''}a^4(\eta'')\\
&&\times\int\frac{d^3{\vec q}}{(2\pi)^3}\frac{d^3{\vec p}}{(2\pi)^3}\frac{d^3{\vec \ell}}{(2\pi)^3}{e^{i(\vec\ell+\vec p+\vec q)\cdot(\vec x-\vec y)}}G_k^C(\eta_1,\eta')\nonumber
\\&&\times G_k^C(\eta_2,\eta'')G_{\ell}^>(\eta',\eta'')G_p^>(\eta',\eta'')G_q^>(\eta',\eta'')\bigg\}\;,
\end{eqnarray*}
where on the seventh line $\ell=\vert\vec k+\vec p+\vec q\rvert$, and on the last two lines $k=\vert\vec\ell+\vec p+\vec q\rvert$. Because in the long-wavelength limit $G^C$-s are momentum-independent, the momentum integrals decouple in this limit and one obtains    
\begin{eqnarray}
&&\!\!\!\!\langle\phi(\vec x,t_1)\phi(\vec y,t_2)\rangle_{\lambda^2,L}^{(3)}={\lambda^2H^2\over12}(\eta_1\eta_2)^{u}\nonumber\\&&~~\times\Bigg\{{3\over r^{2u}}\bigg[\int_{-\infty}^{\eta_1}{d\eta'\over\eta'}(-\eta')^{2u}\int_{-\infty}^{\eta'}{d\eta''\over\eta''}(-\eta'')^{2u}\nonumber\\
&&\quad\quad\quad\quad\quad+\int_{-\infty}^{\eta_2}{d\eta'\over\eta'}(-\eta')^{2u}\int_{-\infty}^{\eta'}{d\eta''\over\eta''}(-\eta'')^{2u}\bigg]\nonumber \\
&&\quad\quad\quad\times\int_0^{-\epsilon r/\eta_{\rm m}}\!\!{\sin w\over w^{2-2u}}{dw}\bigg(\int_0^{-\epsilon/\eta''}{dp\over p^{1-2u}}\bigg)^2\nonumber\\
&&\quad\quad\quad+{1\over r^{6u}}\int_{-\infty}^{\eta_1}{d\eta'\over\eta'}(-\eta')^{2u}\int_{-\infty}^{\eta_2}{d\eta''\over\eta''}(-\eta'')^{2u}\nonumber \\
&&\quad\quad\quad\quad\times\bigg(\int_0^{-\epsilon r/\eta'_{\rm m}}{\sin w\over w^{2-2u}}{dw}\bigg)^3\Bigg\}\;,
\label{2loopSun}
\end{eqnarray}
where  $\eta_{\rm m}$ is the earliest time that accompanies the momentum $\vec k$: $\eta_{\rm m}\equiv {\rm min}(\eta_1,\eta_2,\eta',\eta'')$, and $\eta'_{\rm m}$ is the earliest time that accompanies the momenta $\vec\ell$, $\vec p$ and $\vec q$: $\eta'_{\rm m}\equiv {\rm min}(\eta',\eta'')$. For $Z>1-(2\epsilon^{2})^{-1}$, evaluating \eqref{2loopSun} gives
\begin{eqnarray}
&&\!\!\!\!\!\!\!\!\langle\phi(\vec x,t_1)\phi(\vec y,t_2)\rangle_{\lambda^2,L}^{(3)}\nonumber \\
&=&{\lambda^2H^2\over1024\pi^6u^5}\bigg(1+2uH\vert t_1-t_2\rvert\bigg)e^{-uH\vert t_1-t_2\rvert}\nonumber \\
&&+{\lambda^2H^2\over3072\pi^6u^5}e^{-3uH\vert t_1-t_2\rvert}\;.
\label{2loopSunT}
\end{eqnarray}
For coinciding spacetime points, \eqref{2loopSunT} reduces to
\begin{eqnarray}
\langle\phi^2(\vec x,t)\rangle_{\lambda^2,L}^{(3)}={81\lambda^2H^{12}\over256\pi^6m^{10}}\;.
\label{ssc}
\end{eqnarray}
In the opposite regime, $Z<1-(2\epsilon^{2})^{-1}$, Eq.~\eqref{2loopSun} yields 
\begin{eqnarray}
&&\!\!\!\!\!\!\!\!\langle\phi(\vec x,t_1)\phi(\vec y,t_2)\rangle_{\lambda^2,L}^{(3)}\nonumber \\&=&{\lambda^2H^2\over1024\pi^6u^5}\left(r^2H^2e^{H(t_1+t_2)}\right)^{-u}\nonumber\\&&\quad\quad\quad~\times\left\{1+2u\ln\left({r^2H^2e^{H(t_1+t_2)}}\right)\right\}\nonumber\\&&+{\lambda^2H^2\over3072\pi^6u^5}\left(r^2H^2e^{H(t_1+t_2)}\right)^{-3u}\;.
\end{eqnarray}

Could the ultraviolet divergences and renormalization  change the results obtained in this section? To answer this question, let us look, for example, at one of the terms in \eqref{sunset} rewritten in coordinate space:   
\begin{eqnarray*}
&&\!\!\!\!\int_{-\infty}^{\eta_2}\!{d\eta'}a^4(\eta')\!\int_{-\infty}^{\eta'}\!{d\eta''}a^4(\eta'')\!\int\!d^3\vec{v}\!\!\int\!d^3\vec{w}\,G^{C}(\vec y,t_2;\vec v,t')\nonumber\\&&\times G^{>}(\vec x,t_1;\vec w,t'')G^{C}(\vec v,t';\vec w,t'')\Big[G^{>}(\vec v,t';\vec w,t'')\Big]^2\;.
\end{eqnarray*}
Ultraviolet divergences in this expression occur when $(\vec v,t')=(\vec w,t'')$. Recall that $G^{>}\equiv\langle\phi(\vec{v},t')\phi(\vec{w},t'')\rangle_{\lambda^0}$; hence, around $Z=1$ it is given by \eqref{UV} for spacelike related points and by  
\begin{eqnarray*}
G^{>}(\vec v,t';\vec w,t'')\to{H^2\over8\pi^2}\bigg({1\over u}&-&2-{1\over Z-1\pm i\epsilon}\nonumber\\&-&\ln{(Z-1)\over2}\pm i\pi+\mathcal{O}(u)\bigg)
\end{eqnarray*}
for timelike related points. As shown in Sec.~3.1, the first term in this expression arises solely from modes with small physical momenta.

To write the function $G^C\equiv G^>-G^<=2\,{\rm Im}(G^>)$, we can use the identity   
\begin{eqnarray*}
{1\over x\pm i\epsilon}=P{1\over x}\mp i\pi\delta(x)\;,
\end{eqnarray*}
where $P$ means the principal value. For timelike related points,
\begin{eqnarray*}
G^{C}(\vec v,t';\vec w,t'')\to{H^2\over4\pi^2}\Big(\pm i\pi\delta(Z-1)\pm i\pi\Big)\;,
\end{eqnarray*}
and for spacelike related points it is equal to zero. 

In the product $G^{C}(\vec v,t';\vec w,t'')\Big[G^{>}(\vec v,t';\vec w,t'')\Big]^2$, the leading term in $u$ that does not produce UV divergences is of the order $H^6/u^2$ and derives entirely from the long-wavelength modes. On the other hand, the leading term containing UV divergences is of the order $H^6/u$. Therefore, the renormalization does not affect the value of the two-point function at leading order in small mass.  

Now, taking this into account and combining  \eqref{coinc}, \eqref{coinc-oneloop}, \eqref{2lc}, \eqref{smc} and \eqref{ssc}, we   
get the final expression for the average value of $\langle \phi^2 \rangle$ at coinciding spacetime points in the equilibrium Bunch-Davies state:
\begin{equation}
\langle\phi^2\rangle={3H^4\over8\pi^2m^2}-{27\lambda H^8\over64\pi^4m^6}+{81\lambda^2H^{12}\over64\pi^6m^{10}}+\mathcal O(\lambda^3)\;.
\label{final}
\end{equation}
We have not obtained any instability at the two-loop level conjectured in~\cite{Polyakov:2012uc}. This result is in qualitative agreement with general considerations in~\cite{Marolf:2010nz,Hollands:2010pr,Gautier:2015pca}. Note, however, that the constancy of $\langle\phi^2\rangle$ does not mean that there is no particle creation in this state. Although the notion of particles is not well defined in the long-wavelength (super-Hubble, anti-WKB) regime and $\langle\phi^2\rangle$ includes contributions from both created particles and vacuum polarization, from the fact that the effective number of particles in each Fourier mode 
${\vec k}$ grows with time in the super-Hubble regime, one may say that strong particle creation occurs in the Bunch-Davies state, but this growth is exactly compensated by the universe expansion. This also shows that it is misleading to call the Bunch-Davies state a vacuum, at least if $m^2\ll H^2$. However, it is certainly the equilibrium state.

Our expression \eqref{final} agrees with  the results  in \cite{Gat-Ser,Garb-Rig-Zhu}, which were also obtained by quantum field theory calculation.

\section{Comparison with the Hartree-Fock approximation}
Let us follow paper \cite{Star-Yoko} and write the equation of motion for the scalar field with the action \eqref{action1},
\begin{eqnarray*}
\Box\phi=-m^2\phi-\lambda\phi^3\;,
\end{eqnarray*}
where $\Box\equiv g^{\mu\nu}\nabla_{\mu}\nabla_{\nu}$ is the covariant d'Alembertian.
Multiplying both sides of this equation by $\phi$, using the identity 
\begin{eqnarray*}
\phi\Box\phi={1\over2}\Box{\phi^2}-\partial_{\mu}\phi\partial^{\mu}\phi\;,
\end{eqnarray*}
and taking expectation values of the field operators results in the following equation
\begin{equation*}
\frac12\Box\langle\phi^2\rangle-\langle \partial_{\mu}\phi\,\partial^{\mu}\phi\rangle=-m^2\langle\phi^2\rangle-\lambda \langle \phi^4 \rangle\;.
\end{equation*}
If we use the Hartree-Fock (Gaussian) approximation, $\langle \phi^4 \rangle = 3\langle \phi^2 \rangle^2$, this equation  can be rewritten as
\begin{equation}
\frac12\Box\langle\phi^2\rangle-\langle \partial_{\mu}\phi\,\partial^{\mu}\phi\rangle=-m^2\langle\phi^2\rangle-3\lambda\langle \phi^2 \rangle^2\;.
\label{HF}
\end{equation}

In the case of a massless noninteracting field, the long-wavelength part of $\langle\phi^2\rangle$ is given by   
\begin{eqnarray}
\langle\phi^2(\vec x,t)\rangle_{L}&=&\int^{-\epsilon/\eta}\frac{d^3{\vec k}}{(2\pi)^3}\phi_k(\eta)\phi_k^*(\eta)={H^2\over4\pi^2}\int_{\kappa}^{-\epsilon/\eta}{dk\over k}\nonumber\\&=&-{H^2\over4\pi^2}\ln\left({-\kappa\eta\over\epsilon}\right)\;,
\label{msl}
\end{eqnarray}
where Eq.~\eqref{longmodes} with $u=0$ was used to express the mode functions. We introduced an infrared cutoff $\kappa$ for the comoving momentum $k$ since the integral is divergent at $k=0$. In terms of cosmic time \eqref{msl} becomes
\begin{eqnarray}
\langle\phi^2\rangle_{L}={H^3\over4\pi^2}(t-t_0)\;,
\label{lg}
\end{eqnarray}  
where $t_0\equiv(1/H)\ln(\kappa/H\epsilon)$; thus, it grows linearly with time \cite{Allen:1985ux,Allen:1987tz,Starobinsky:1982ee,Linde:1982uu,Vilenkin:1982wt}. This means that the only term that survives in the d'Alembertian operator is the first time derivative: 
\begin{eqnarray}
\Box\langle\phi^2\rangle_{L}=3H{\partial\over\partial t}\langle\phi^2\rangle_{L}\;.
\label{dtd}
\end{eqnarray}
The second term on the left side of Eq.~\eqref{HF} contains two derivatives and, therefore, is a constant proportional to $\epsilon^2$,
\begin{equation}
\langle \partial_{\mu}\phi\,\partial^{\mu}\phi\rangle_{L}\propto\epsilon^2\;;
\label{2der}
\end{equation} 
hence, it can be neglected.

When $\lambda$ and $m$ are small, expressions \eqref{dtd} and \eqref{2der} are still approximately correct, so Eq.~\eqref{HF} reduces to
\begin{equation}
{\partial\over\partial t}\langle\phi^2\rangle_{L}=\frac{H^3}{4\pi^2}-{2m^2\over3H}\langle \phi^2 \rangle_L-\frac{2\lambda}{H}\langle \phi^2 \rangle_L^2\;,
\end{equation}
where the first term on the right side is needed to recover \eqref{lg} in the limit of zero mass and coupling. As $t\to\infty$, all of the solutions to this equation approach an equilibrium value that satisfies      
\begin{equation}
\frac{H^3}{4\pi^2}-{2m^2\over3H}\langle \phi^2 \rangle_L-\frac{2\lambda}{H}\langle \phi^2 \rangle_L^2=0\;.
\label{shf}
\end{equation}
For $\lambda=0$, we have  
\begin{equation}
\langle \phi^2\rangle_L={3H^4\over8\pi^2m^2}\;.
\label{linear}
\end{equation} 
When $\lambda\neq0$, Eq~\eqref{shf} gives
\begin{equation}
\langle \phi^2 \rangle_L={m^2\over6\lambda}\left(\sqrt{1+{9\lambda H^4\over2\pi^2m^4}}-1\right);
\label{root}
\end{equation}
we chose the root that coincides with \eqref{linear} in the limit $\lambda\to0$. Assuming that $\lambda H^4/m^4\ll1$, and expanding \eqref{root} yields  
\begin{equation}
\langle \phi^2\rangle_L={3H^4\over8\pi^2m^2}-{27\lambda H^8\over64\pi^4m^6}+{243\lambda^2H^{12}\over256\pi^6m^{10}}+\mathcal O(\lambda^3)\;.
\label{hfexp}
\end{equation} 
Comparing this expansion with the two-loop results of the previous section (with \eqref{final} in particular), we see that they match at zeroth- and first-order in $\lambda$, but there is a mismatch at second order: the $\lambda^2$-term in \eqref{hfexp} omits the contribution of the sunset diagram and is equal to the sum of \eqref{2lc} and \eqref{smc}. As a result, in the Hartree-Fock approximation this term is $4/3$ times smaller.  Also, it has to be concluded that Eq.~\eqref{root} resumes all cactus type diagrams of the perturbation theory.

\section{Comparison with the stochastic approach}
The stochastic approach argues \cite{Star,Star-Yoko} that the behavior of the long-wavelength part of the quantum field $\phi(\vec x,t)$ in de Sitter spacetime can be modeled by an auxiliary  classical stochastic variable $\varphi$ with a probability distribution $\rho(\varphi,t)$ that satisfies the Fokker-Planck equation   
\begin{equation}
{\partial\rho\over\partial t} 
= {H^3\over8\pi^2}{\partial^2\rho\over\partial\varphi^2} 
+ {1\over3H}{\partial\over\partial\varphi}
\biggl({\partial V\over\partial\varphi} \rho(t,\varphi) \biggr)\;;
\label{F-P}
\end{equation}
that is, the expectation value of any quantity constructed from the long-wavelength part of $\phi(\vec x,t)$ is equal to the expectation value of the same quantity constructed from  the variable $\varphi$. $V(\varphi)$ is a function of the stochastic variable and has the same functional form as the corresponding potential of the quantum field theory: for the theory with the action \eqref{action1} this potential has the form
\begin{equation}
V(\varphi)={1\over2}{m^2}\varphi^2+{\lambda\over4}\varphi^4\;.
\label{pot}
\end{equation}

At late times any solution of the equation \eqref{F-P} approaches the static equilibrium solution
\begin{equation}
\rho_{\rm eq}(\varphi) = N^{-1}\exp\left(-\frac{8\pi^2}{3H^4}V(\varphi)\right),
\label{stat}
\end{equation}
where $N$ is the normalization fixed by the condition 
\begin{eqnarray*}
\int_{-\infty}^{\infty}\rho_{\rm eq}(\varphi)\,d\varphi=1\;.
\end{eqnarray*}
In particular, with the potential \eqref{pot} we can calculate this normalization explicitly
\begin{eqnarray}
N&=&\int_{-\infty}^{\infty}\exp\left[-{8\pi^2\over3H^4}\left({\lambda\varphi^4\over4}+{m^2\varphi^2\over2}\right)\right]d\varphi\nonumber\\&=&{m\over{\sqrt{2\lambda}}}\exp(z)\mathcal{K}_{1\over4}(z)\;,
\label{norm}
\end{eqnarray}
where $\mathcal{K}_{1\over4}(z)$ is a modified Bessel function of the second kind, and $z\equiv{\pi^2m^4\over3\lambda H^4}$. The distribution \eqref{stat} can be used to compute the expectation value of $\varphi^2$:
\begin{eqnarray}
\langle \varphi^2 \rangle&=&\int_{-\infty}^{\infty}\varphi^2 \rho_{\rm eq}(\varphi)d\varphi\nonumber\\&=&N^{-1}\int_{-\infty}^{\infty}\varphi^2\exp\left[-{8\pi^2\over3H^4}\left({\lambda\varphi^4\over4}+{m^2\varphi^2\over2}\right)\right]d\varphi\;.\nonumber\\
\label{two}
\end{eqnarray}
The integral above can be expressed in the following way:
\begin{eqnarray}
&&\!\!\!\!\int_{-\infty}^{\infty}\!\!\varphi^2\exp\left[-{8\pi^2\over3H^4}\left({\lambda\varphi^4\over4}+{m^2\varphi^2\over2}\right)\right]\nonumber \\
&&={\pi m^3\over 4\lambda^{3/2}}\exp(z)\left[{\cal I}_{1\over4}(z)-{\cal I}_{-{1\over4}}(z)+{\cal I}_{5\over4}(z)-{\cal I}_{3\over4}(z)\right]\nonumber \\
&&~~+{3H^4\over8\pi m}{\exp(z)\over\lambda^{1/2}}{\cal I}_{1\over4}(z)\,,
\label{int}
\end{eqnarray}
where $\mathcal{I}_{\nu}(z)$ are modified Bessel functions of the first kind. Combining the right sides of \eqref{norm} and \eqref{int}, and using the connection formula
\begin{eqnarray*}
\mathcal{K}_{\nu}(z)={\pi\over2}{\mathcal{I}_{-\nu}(z)-\mathcal{I}_{\nu}(z)\over\sin(\nu\pi)}
\end{eqnarray*}
and the recurrence relation
\begin{eqnarray*}
\mathcal{I}_{\nu-1}(z)-\mathcal{I}_{\nu+1}(z)={2\nu\over z}\mathcal{I}_{\nu}(z),
\end{eqnarray*}
we obtain a fairly concise expression for \eqref{two}
\begin{eqnarray} 
\langle\varphi^2\rangle={m^2\over2\lambda}{\mathcal{K}_{3\over4}(z)\over\mathcal{K}_{1\over4}(z)}-{m^2\over2\lambda}.
\label{BK}
\end{eqnarray}
Interestingly, this result is in agreement with the expression obtained in \cite{Beneke:2012kn} for the amplitude of the zero mode in Euclidean de Sitter space.

Expanding \eqref{BK} in the limit $\lambda H^4/m^4\ll1$ (which corresponds to $z\gg1$) gives
\begin{equation}
\langle\varphi^2\rangle={3H^4\over8\pi^2m^2}-{27\lambda H^8\over64\pi^4m^6}+{81\lambda^2H^{12}\over64\pi^6m^{10}}
-\frac{24057\lambda^3H^{16}}{4096\pi^8m^{14}}
+\mathcal O(\lambda^4)\;.
\label{third-ord}
\end{equation}
As we can see, this result coincides exactly with the result \eqref{final} of the quantum field theory calculations of Sec.~3, and unlike the Hartree-Fock approximation \eqref{hfexp}, it includes the contribution of the sunset diagram; that is, the $\lambda^2$-term is equal to the sum of \eqref{2lc}, \eqref{smc} and \eqref{ssc}. In Eq. \eqref{third-ord} we have also given the expression for the term proportional to $\lambda^3$. The corresponding expression obtained by the field-theoretical methods will be presented in the future work \cite{future}.

  Note also that in the opposite limit $z\to 0$, the known expression for the massless self-interacting field arises~\cite{Star-Yoko} that would require summation of all loops in the standard field-theoretic approach:
\begin{equation}
\langle\varphi^2\rangle = \sqrt{\frac{3}{2\pi^2}}\, \frac{\Gamma(3/4)}{\Gamma(1/4)}\, \frac{H^2}{\sqrt{\lambda}}
\approx 0.132\, \frac{H^2}{\sqrt{\lambda}}\, .
\label{79}
\end{equation}
In this case, the Hartree-Fock approximation \eqref{root} produces the result that is only 1.17 times smaller than \eqref{79} (see \cite{Star-Yoko}).

Interestingly, by differentiating  the formula \eqref{norm} with respect to $m^2$  and using the recurrence relations for the modified Bessel functions of the second kind  (see e.g. \cite{Grad}),
\begin{equation*}
z\frac{d}{dz}\mathcal{K}_{\nu}(z) +\nu\mathcal{K}_{\nu}(z)=-z\mathcal{K}_{\nu-1}(z),
\end{equation*}
and the fact that $\mathcal{K}_{-\nu}(z)=\mathcal{K}_{\nu}(z)$, 
one can express an arbitrary average value $\langle\varphi^{2n}\rangle$ in terms of the ratio $\mathcal{K}_{3\over 4}(z)/\mathcal{K}_{1\over 4}(z)$ . For example, for $\langle\varphi^{4}\rangle$, we obtain
\begin{equation*}
\langle\varphi^4\rangle = \frac{3H^4}{8\pi^2\lambda}+\frac{m^4}{2\lambda^2}\left(1-\frac{\mathcal{K}_{3\over 4}(z)}{\mathcal{K}_{1\over 4}(z)}\right).
\end{equation*}

As discussed in \cite{Star-Yoko}, the long-wavelength two-point function of $\phi(\vec x,t)$ too can be calculated by using the classical stochastic variable $\varphi$: if the points $(\vec x,t_1)$ and $(\vec y,t_2)$ are timelike or lightlike related, this correlation function is given by      
\begin{equation}
\langle\phi(\vec x,t_1)\phi(\vec y,t_2)\rangle_L=\langle\varphi(t_1)\varphi(t_2)\rangle\;.
\end{equation}
If the correlation function $\langle\varphi(t_1)\varphi(t_2)\rangle$ depends only on the absolute value of the time difference $T\equiv\vert t_1-t_2\rvert$, it can be expressed as \cite{Star-Yoko,Risken} 
\begin{equation}
\langle\varphi(t_1)\varphi(t_2)\rangle=\int_{-\infty}^{\infty}\varphi\,\Xi(\varphi,T)d\varphi\;,
\label{2p}
\end{equation}
where the function $\Xi(\varphi,T)$ satisfies the Fokker-Planck equation \eqref{F-P}, 
\begin{equation}
{\partial\Xi\over\partial T} 
= {H^3\over8\pi^2}{\partial^2\Xi\over\partial\varphi^2} 
+ {1\over3H}{\partial\over\partial\varphi}
\biggl({\partial V\over\partial\varphi} \Xi(\varphi,T) \biggr)\;,
\label{F-P2}
\end{equation}
with the initial condition 
\begin{equation}
\Xi(\varphi,0)=\varphi\rho_{\rm eq}(\varphi)\;,
\label{inc}
\end{equation} 
and $\rho_{\rm eq}(\varphi)$ is the equilibrium solution \eqref{stat}. Derivatives of $\langle\varphi(t_1)\varphi(t_2)\rangle$ at $T=0$ can be computed by using Eqs.~\eqref{2p}--\eqref{inc}: 
\begin{equation}
{\partial\over\partial T}\langle\varphi(t_1)\varphi(t_2)\rangle\bigg|_{T=0}=-{H^3\over8\pi^2}\;,
\end{equation}
\begin{equation}
{\partial^2\over\partial T^2}\langle\varphi(t_1)\varphi(t_2)\rangle\bigg|_{T=0}={H^2\over24\pi^2}\Big(3\lambda\langle\varphi^2\rangle+m^2\Big)\;,
\end{equation}
and so on. It is easy to confirm that the $T$-derivatives of the two-point correlation function computed in Sec.~3 (for $Z>1-{1\over2\epsilon^2}$ case) satisfy these equalities as well.
\section{Exponentiation of the perturbative series}
Let us put together the calculations of Sec.~3: up to two loops, for $Z>1-(2\epsilon^{2})^{-1}$ the long-wavelentgth two-point correlation function is given by  
\begin{eqnarray*}
&&\!\!\!\!\!\!\!\!\langle\phi(\vec x,t_1)\phi(\vec y,t_2)\rangle_{L}\\
&=&{H^2\over8\pi^2u}e^{-uHT}-{\lambda H^2\over64\pi^4u^3}\bigg(1+uHT\bigg)e^{-uHT}\\
&&+{\lambda^2H^2\over512\pi^6u^5}\bigg(1+uHT+{1\over2}u^2H^2{T}^2\bigg)e^{-uHT}\nonumber\\
&&+{\lambda^2H^2\over512\pi^6u^5}\bigg(1+uHT\bigg)e^{-uHT}\\
&&+{\lambda^2H^2\over1024\pi^6u^5}\bigg(1+2uHT\bigg)e^{-uHT}\\
&&+{\lambda^2H^2\over3072\pi^6u^5}e^{-3uHT}+\mathcal O(\lambda^3)\;,
\end{eqnarray*}
where $T\equiv\vert t_1-t_2\rvert$. It will be convenient to reorganize this expression in the following way:
\begin{eqnarray}
&&\!\!\!\!\!\!\!\!\langle\phi(\vec x,t_1)\phi(\vec y,t_2)\rangle_{L}\nonumber \\
&=&{H^2\over8\pi^2u}\left(1-{\lambda\over8\pi^2u^2}+{5\lambda^2\over128\pi^4u^4}\right)e^{-uHT}\nonumber \\
&&-{\lambda H^3T\over64\pi^4u^2}\left(1-{3\lambda\over8\pi^2u^2}\right)e^{-uHT}\nonumber\\&&+{\lambda^2H^4{T}^2\over1024\pi^6u^3}e^{-uHT}+{\lambda^2H^2\over3072\pi^6u^5}e^{-3uHT}+\mathcal O(\lambda^3)\nonumber\\&=&{H^2\over8\pi^2u}\left(1-{\lambda\over8\pi^2u^2}+{5\lambda^2\over128\pi^4u^4}+\mathcal O(\lambda^3)\right)\nonumber \\
&&\times\left[1-C_1{\lambda HT\over8\pi^2u}+{C_2\over2}\left({\lambda HT\over8\pi^2u}\right)^2+\mathcal O(\lambda^3)\right]e^{-uHT}\nonumber\\&&+{\lambda^2H^2\over3072\pi^6u^5}e^{-3uHT}+\dots\;,
\label{series}
\end{eqnarray}
where $C_1\equiv(1-{3\lambda\over8\pi^2u^2})(1-{\lambda\over8\pi^2u^2}+{5\lambda^2\over128\pi^4u^4})^{-1}$ and $C_2\equiv(1-{\lambda\over8\pi^2u^2}+{5\lambda^2\over128\pi^4u^4})^{-1}$. Upon expanding $C_1$ and $C_2$ in powers of $\lambda$, the second term in the square brackets of \eqref{series} becomes
\begin{eqnarray*}
-C_1{\lambda HT\over8\pi^2u}=-{\lambda HT\over8\pi^2u}+{\lambda^2HT\over32\pi^4u^3}+\mathcal O(\lambda^3)\;,
\label{second}
\end{eqnarray*}
and the third term
\begin{eqnarray*}
{C_2\over2}\left({\lambda HT\over8\pi^2u}\right)^2={1\over2}\left({\lambda HT\over8\pi^2u}\right)^2+\mathcal O(\lambda^3).
\label{third}
\end{eqnarray*}
Hence the expression in square brackets reduces to 
\begin{equation*}
1-{\lambda HT\over8\pi^2u}+{\lambda^2HT\over32\pi^4u^3}+{1\over2}\left({\lambda HT\over8\pi^2u}\right)^2+\mathcal O(\lambda^3)\;.
\end{equation*}
To second order in $\lambda$, this matches with the first three terms in the Taylor series of the exponential function 
\begin{eqnarray}
\exp\left[{-{\lambda HT\over8\pi^2u}\left(1-{\lambda\over4\pi^2u^2}\right)}\right],
\end{eqnarray}
so it is plausible that an infinite series of diagrams may be resummed into this exponent. With this assumption, we arrive at   
\begin{eqnarray}
&&\!\!\!\!\!\!\!\!\langle\phi(\vec x,t_1)\phi(\vec y,t_2)\rangle_{L}\nonumber \\
&=&{H^2\over8\pi^2u}\left(1-{\lambda\over8\pi^2u^2}+{5\lambda^2\over128\pi^4u^4}+\mathcal O(\lambda^3)\right)\nonumber \\
&&\times\exp\bigg[{-uHT\left(1+{\lambda\over8\pi^2u^2}-{\lambda^2\over32\pi^4u^4}+\mathcal O(\lambda^3)\right)}\bigg]\nonumber\\&&+{\lambda^2H^2\over3072\pi^6u^5}\exp\Big[{-3uHT}\Big]+\dots\;.
\label{expt}
\end{eqnarray}
Similarly, for $Z<1-(2\epsilon^{2})^{-1}$ we obtain
\begin{eqnarray}
&&\!\!\!\!\!\!\!\!\langle\phi(\vec x,t_1)\phi(\vec y,t_2)\rangle_{L}\nonumber \\
&=&{H^2\over8\pi^2u}\left(1-{\lambda\over8\pi^2u^2}+{5\lambda^2\over128\pi^4u^4}+\mathcal O(\lambda^3)\right)\nonumber \\
&&\times\left(r^2H^2e^{H(t_1+t_2)}\right)^{-u\left(1+{\lambda\over8\pi^2u^2}-{\lambda^2\over32\pi^4u^4}+\mathcal O(\lambda^3)\right)}\nonumber\\&&+{\lambda^2H^2\over3072\pi^6u^5}\left(r^2H^2e^{H(t_1+t_2)}\right)^{-3u}+\dots\;,
\label{exps}
\end{eqnarray}
and it is evident that in this regime the equal-time correlation function depends only on the physical spatial distance $R\equiv re^{Ht}$.

From the expressions \eqref{expt} and \eqref{exps}, we see that the perturbative corrections don't change the long-wavelength part of the commutator: just as in the free theory case, it is equal to zero both for timelike and spacelike related points.

When $|Z|\gg1$, both \eqref{expt} and \eqref{exps} can be expressed in a de Sitter invariant form. As was noted in Sec.~3.1, for large timelike separations $2Z\approx e^{HT}$, and for large spacelike separations $2Z\approx-H^2r^2e^{H(t_1+t_2)}$. Therefore,
\begin{eqnarray}
&&\!\!\!\!\!\!\!\!\langle\phi(\vec x,t_1)\phi(\vec y,t_2)\rangle_{L}\nonumber \\
&=&{H^2\over8\pi^2u}\left(1-{\lambda\over8\pi^2u^2}+{5\lambda^2\over128\pi^4u^4}+\mathcal O(\lambda^3)\right)\nonumber \\
&&\times|Z|^{-u\left(1+{\lambda\over8\pi^2u^2}-{\lambda^2\over32\pi^4u^4}+\mathcal O(\lambda^3)\right)}\nonumber\\&&+{\lambda^2H^2\over3072\pi^6u^5}|Z|^{-3u}+\dots\;
\label{expz}
\end{eqnarray}
both for large timelike and large spacelike separations. This expression is also correct for coinciding spacetime points, since in this case $Z=1$ and \eqref{expz} gives the same result as \eqref{expt} with $T$ set equal to zero. In \cite{Gat-Ser}, momentum representation of the unequal-time correlation function is calculated by using the Schwinger-Dyson equations to resum the infinite series of self-energy insertions. A comparison of its results with \eqref{expt}--\eqref{expz} is presented in Appendix B.

As was shown in \cite{Star-Yoko}, correlation functions deduced by the stochastic approach have a form of an infinite sum of exponentials; e.g., if points are timelike related, the two-point function is given by  
\begin{equation*}
\langle\phi(\vec x,t_1)\phi(\vec y,t_2)\rangle_L=\langle\varphi(t_1)\varphi(t_2)\rangle=N\sum_{n=1}^{\infty}A_n^2e^{-\Lambda_n T},
\end{equation*}
where $N$ is the normalization constant \eqref{norm}. By comparing this with Eq.~\eqref{expt}, we can identify the first three coefficients $A_n$ and $\Lambda_n$:   
\begin{eqnarray*}
NA_1^2&=&{H^2\over8\pi^2u}\left(1-{\lambda\over8\pi^2u^2}+{5\lambda^2\over128\pi^4u^4}+\mathcal O(\lambda^3)\right)\;,\nonumber\\
A_2&=&0\,,\,NA_3^2={\lambda^2H^2\over3072\pi^6u^5}+\mathcal O(\lambda^3)\;,\nonumber\\
\Lambda_1&=&\left(1+{\lambda\over8\pi^2u^2}-{\lambda^2\over32\pi^4u^4}+\mathcal O(\lambda^3)\right)uH\;,\nonumber\\
\Lambda_3&=&3uH+\mathcal O(\lambda)\;.  
\end{eqnarray*}
These coefficients coincide with the results of \cite{Mark-et,Mor-Ser} where they are derived using some versions of the stochastic approach (e.g., \cite{Mor-Ser} uses the formulation of the stochastic theory in terms of a supersymmetric one-dimensional field theory).

As $T\to\infty$, the two-point function \eqref{expt} decays with the characteristic correlation time   
\begin{equation*}
T_c\sim{1\over uH}={3H\over m^2}\gg{1\over H}\;.    
\end{equation*}
Similarly, it follows from \eqref{exps} that as $R\to\infty$, the equal-time correlation function decays with the characteristic correlation length
\begin{equation*} 
R_c\sim{1\over H}\exp\left({3H^2\over2m^2}\right)\;.    
\end{equation*}
The behavior of \eqref{exps} differs from a much faster exponential decay of the equal-time correlation function in flat spacetime: $\langle\phi(\vec x,t)\phi(\vec y,t)\rangle_{\rm flat}\sim\sqrt{m/r^3}e^{-mr}$ as $r\to\infty$.  

\section{Conclusions}
In Minkowski spacetime, quantum fluctuations decay quickly with scale and are significant only at short distances. The situation is different in de Sitter spacetime. Here the exponential expansion stretches short-wavelength modes to super-Hubble scales without decreasing its amplitude too much. On super-Hubble scales, the amplitude of fluctuations is scale-independent for the massless minimally coupled scalar field, and only weakly decays with scale when $0<m^2\ll H^2$.                  

In this article we have calculated---up to two loops---the long-wavelength two-point function for a scalar theory with a small mass and a quartic interaction. It has been shown that it is de Sitter invariant for coinciding points as well as at large spacelike and large timelike separations. We have also demonstrated that the commutator of the long-wavelength part of the field is equal to zero both at the free theory level and with the perturbative corrections.    

Our results are in agreement with Starobinsky's stochastic approach in which the coarse-grained quantum field is equivalent to a classical stochastic quantity.

\begin{acknowledgements}
The work of A.~K. and A.~S. was partially supported by the RFBR Grant No. 20-02-00411. The work of T. V. was partially supported by the research grant ``The
Dark Universe: A Synergic Multimessenger Approach'' No. 2017X7X85K under the program PRIN 2017 funded
by the Ministero dell'Istruzione, delle Universit\`a e della Ricerca (MIUR).
\end{acknowledgements}

\appendix 
\section{Geodesics in de Sitter spacetime}
De Sitter spacetime can be represented as the four-dimensional hyperboloid
\begin{equation}
(X^1)^2+(X^2)^2+(X^3)^2+(X^4)^2-(X^0)^2=H^{-2}
\label{Z3}
\end{equation}
embedded in five-dimensional Minkowski spacetime. In this appendix we set $H=1$.

For two points in de Sitter spacetime, the only invariant quantity associated with these points is the function
\begin{equation}
Z(X,Y) \equiv X^1Y^1+X^2Y^2+X^3Y^3+X^4Y^4-X^0Y^0\;.
\label{Z}
\end{equation}
Let us consider the spacetime interval between these points:
\begin{eqnarray}
s^2(X,Y)&\equiv&(X^0-Y^0)^2-(X^1-Y^1)^2-(X^2-Y^2)^2\nonumber \\
&&-(X^3-Y^3)^2-(X^4-Y^4)^2.
\label{Z2}
\end{eqnarray}
Using the hyperboloid equation \eqref{Z3} (with $H=1$) and the definition of $Z$ from Eq.~\eqref{Z}, we obtain
\begin{equation}
Z(X,Y)=1+{1\over2}s^2(X,Y).
\label{Z4}
\end{equation}
Thus, if the interval between $X$ and $Y$ is timelike, then $Z(X,Y) > 1$; if the interval is spacelike, then $Z(X,Y) < 1$, and if the interval is lightlike, then $Z(X,Y)=1$. Note the if $\overline{Y}$ is the antipodal point of $Y$ (that is, $\overline{Y} = -Y$), then $Z(X,\overline{Y}) = -Z(X,Y)$; hence, if $Z(X,Y) < -1$, the interval between $X$ and $\overline{Y}$ is timelike.    

Let us show that the geodesics in de Sitter spacetime are intersections of the hyperboloid with hyperplanes passing through the origin of the ambient spacetime just like the geodesics on the sphere are the great circles. To simplify notations, we will consider two-dimensional de Sitter spacetime embedded into 3-dimensional flat spacetime. In terms of the spherical coordinates
\begin{eqnarray}
&&X^0=\sinh t,\nonumber \\
&&X^1=\cosh t \cos\phi,\nonumber \\
&&X^2=\cosh t\sin\phi\;,
\label{Z5}
\end{eqnarray}
which  cover all of the de Sitter hyperboloid, the metric has the form 
\begin{equation}
ds^2 = dt^2 -\cosh^2t d\phi^2\;.
\label{Z6}
\end{equation} 
The Christoffel symbols for this metric are 
\begin{equation}
\Gamma_{\phi t}^{\phi} = \tanh t,\ \ \Gamma_{\phi\phi}^t = \sinh t\cosh t\;.
\label{Z7}
\end{equation} 
The equations for the geodesics are 
\begin{equation}
\frac{d^2\phi}{d\tau^2}+2\Gamma_{\phi t}^{\phi}\frac{d\phi}{d\tau}\frac{dt}{d\tau}=0\;,
\label{geod}
\end{equation}
\begin{equation}
\frac{d^2 t}{d\tau^2}+\Gamma_{\phi\phi}^t\left(\frac{d\phi}{d\tau}\right)^2=0\;.
\label{geod1}
\end{equation}
Upon substituting the expressions for the Christoffel symbols \eqref{Z7} into Eqs.~\eqref{geod}-\eqref{geod1} , we obtain
\begin{equation}
\frac{d^2\phi}{d\tau^2}+2\tanh t\frac{d\phi}{d\tau}\frac{dt}{d\tau}=0\;,
\label{geod2}
\end{equation}
\begin{equation}
\frac{d^2 t}{d\tau^2}+\sinh t\cosh t\left(\frac{d\phi}{d\tau}\right)^2=0\;.
\label{geod3}
\end{equation}
Equation \eqref{geod2} gives 
\begin{equation}
\frac{d\phi}{d\tau} = \frac{A}{\cosh^2 t}\;,
\label{geod4}
\end{equation}
where $A$ is an integration constant. Substituting the expression \eqref{geod4} into Eq.~\eqref{geod3} yields
\begin{equation}
\frac{d^2 t}{d\tau^2} = -\frac{A^2\sinh t}{\cosh^3 t}\;.
\label{geod5}
\end{equation}
Multiplying Eq.~\eqref{geod5} by $2(dt/d\tau)$ produces 
\begin{equation}
\frac{d}{d\tau}\left(\frac{dt}{d\tau}\right)^2 = A^2\frac{d}{d\tau}\left(\frac{1}{\cosh^2 t}\right).
\label{geod6}
\end{equation}
After integrating Eq.~\eqref{geod6}, we obtain 
\begin{equation}
\frac{dt}{d\tau}=\sqrt{\frac{A^2}{\cosh^2t}+B},
\label{geod7}
\end{equation}
where $B$ is an integration constant.  Comparing Eqs.~\eqref{geod4} and \eqref{geod7} gives the expression for the geodesics in terms of the variables $\phi$ and $t$:
\begin{equation}
\frac{d\phi}{dt} = \frac{A}{\cosh t\sqrt{A^2+B\cosh^2t}}\;.
\label{geod8}
\end{equation}

The equation for a plane passing through the origin of the ambient spacetime can be written as 
\begin{equation}
\alpha X_0 -\beta X_1-\gamma X_2\;.
\label{geod9}
\end{equation}
To obtain the expression for the intersection of \eqref{geod9} with the de Sitter hyperboloid, one should substitute the expressions \eqref{Z5} into Eq. \eqref{geod9}:
\begin{equation}
\alpha \sinh t -\beta \cosh t\cos\phi-\gamma \cosh t\sin\phi=0.
\label{geod10}
\end{equation}
We can rewrite Eq.~\eqref{geod10} as follows:
\begin{equation}
\alpha \tanh t = \sqrt{\alpha^2+\beta^2}\sin(\phi+\phi_0)\;. 
\label{geod11}
\end{equation}
Eq. \eqref{geod11} gives the differential equation
\begin{equation}
\frac{\alpha}{\cosh^2t}dt = \sqrt{\alpha^2+\beta^2}\cos(\phi+\phi_0)d\phi\;.
\label{geod12}
\end{equation}
From Eq.~\eqref{geod11} we find that 
\begin{equation}
\cos(\phi+\phi_0)=\sqrt{1-\frac{\alpha^2}{\beta^2+\gamma^2}\tanh^2t}\;,
\label{geod13}
\end{equation}  
and using Eqs.~\eqref{geod12} and \eqref{geod13}, we obtain
\begin{equation}
\frac{d\phi}{dt} = \frac{\alpha}{\cosh t\sqrt{(\beta^2+\gamma^2-\alpha^2)\cosh^2t+\alpha^2}}\;.
\label{geod14}
\end{equation}
The equation \eqref{geod14} coincides with Eq.~\eqref{geod8}, and hence we have shown that intersections of planes through the origin with the de Sitter hyperboloid are geodesics. 

This result should not depend on the choice of the coordinate system. However, for completeness we will reproduce it in the flat coordinates
\begin{eqnarray}
&&X^0 = \sinh t +\frac12x^2e^t,\nonumber \\
&&X^1 = xe^t,\nonumber \\
&&X^2 =\cosh t - \frac12x^2e^t\;.
\label{geod15}
\end{eqnarray}
In these coordinates the metric is 
\begin{equation}
ds^2 = dt^2-e^{2t}dx^2\;,
\label{geod16}
\end{equation}
and the equations for the geodesics are 
\begin{equation}
\frac{d^2x}{d\tau^2}+2\frac{dx}{d\tau}\frac{d\phi}{d\tau}=0\;,
\label{geod17}
\end{equation}
\begin{equation}
\frac{d^2t}{d\tau^2}+e^{2t}\left(\frac{dx}{d\tau}\right)^2\;.
\label{geod18}
\end{equation}
Integrating Eq. \eqref{geod17} gives
\begin{equation}
\frac{dx}{d\tau} = Ce^{-2t}\;,
\label{geod19}
\end{equation} 
where $C$ is an integration constant. 
Substituting the expression \eqref{geod19} into Eq. \eqref{geod18} and proceeding in a manner similar to that used for the spherical coordinates yields
\begin{equation}
\frac{dt}{d\tau}=\sqrt{C^2e^{-2t}+D}\;.
\label{geod20}
\end{equation}
Then, 
\begin{equation}
\frac{dx}{dt} = \frac{Ce^{-2t}}{\sqrt{C^2e^{-2t}+D}}\;.
\label{geod21}
\end{equation}
This equation can be integrated explicitly:
\begin{equation}
x(t) = F - \frac{1}{C}\sqrt{D+C^2e^{-2t}}\;.
\label{geod22}
\end{equation}
In terms of the coordinates \eqref{geod15}, intersections of planes through the origin of the ambient spacetime with the de Sitter hyperboloid are given by
\begin{eqnarray}
&&\delta\left(\sinh t+\frac12x^2e^t\right)+\varepsilon x e^t+\eta\left(\cosh t-\frac12x^2e^t\right)=0\;.\nonumber \\
\label{geod23}
\end{eqnarray}
Solving this quadratic equation with respect to $x$, we find 
\begin{equation}
x(t) = -\frac{\varepsilon}{\delta-\eta}+\sqrt{\frac{\varepsilon^2-\delta^2+\eta^2}{(\delta-\eta)^2}+e^{-2t}}.
\label{geod24}
\end{equation}
The form of Eq. \eqref{geod24} coincides with Eq. \eqref{geod22}.

For any two points in Minkowski spacetime, there always exist a geodesic that connects these points. This geodesics can be timelike, lightlike or spacelike, but it always exists. In de Sitter spacetime the situation is different. Let us take two arbitrary points $X$ and $Y$ on the de Sitter hyperboloid. When is it possible to connect them by a geodesic? 

To answer this question, consider the plane that passes through these two points and the origin of the coordinates. Note that if some point $U$ belongs to the intersection of this plane with the de Sitter hyperboloid, then its antipodal point 
$\overline{U} = - U$ also belongs to this intersection. If the angle between this plane and the plane $X^0 = 0$ is less than $\pi/4$, its intersection with the hyperboloid is an ellipse. In this case, all vectors tangent to this curve are spacelike, and we can call this geodesic spacelike. Therefore, $Z(X,Y)<1$ for any two points on the ellipse, and since their antipodal points are also on this ellipse, $Z(X,{\overline Y})<1$ as well. Hence, any two points that can be connected by an elliptical geodesic should satisfy the condition that $-1<Z(X,Y)<1$.

If the angle between a plane through a pair of points and the origin and the plane $X^0 = 0$ is $\geq\pi/4$, its intersection with the hyperboloid gives a hyperbola with two branches. Any two points on the same branch are timelike related (or lightlike related if the angle is equal to $\pi/4$), which means that $Z(X,Y)\geq1$. The antipodal point of any point on a given branch is located on the other branch, and hence for any two points lying on different branches $Z(X,Y)\leq-1$. The latter points, obviously, cannot be not connected by any geodesic.  

\section{A comparison with the results in $p$-representation} 
For completeness of presentation and for convenience of readers, in this Appendix we will present a detailed comparison of our results \eqref{expt}--\eqref{expz} with the two-point function calculated in \cite{Gat-Ser} in the framework of momentum representation \cite{Par-Ser}. We will use formulas from \cite{Gat-Ser} and considerations of Appendix B of \cite{Gautier:2015pca} concerning the relation between Green's functions in $p$-representation and in coordinate space.

In $d+1$ dimensional flat de Sitter space, where the Hubble parameter is set to be equal to 1, the metric is 
\begin{equation}
ds^2 = \eta^{-2}(-d\eta^2+d{\bf X}\cdot d{\bf X}).
\label{GS}
\end{equation}
 A Green's function can be represented as follows:
 \begin{equation}
 G(x,x') = \int\frac{d^dK}{(2\pi)^d}e^{i{\bf K}\cdot({\bf X}-{\bf X'})}\tilde{G}(\eta,\eta',K).
 \label{GS1}
 \end{equation}
 De Sitter symmetries ensure that the correlation function has the following $p$-representation \cite{Par-Ser}:
 \begin{equation}
 \tilde{G} (\eta,\eta',K) = \frac{(\eta\eta')^{\frac{d-1}{2}}}{K}\hat{G}(p,p'),
 \label{GS2}
 \end{equation}
 where 
 \begin{equation*}
 p = -K\eta,\ \ p'=-K\eta'.
 \end{equation*}
 Now, let us suppose that for the Green's function which we would like to calculate the expression $\hat{G}$ is proportional to
\begin{equation}
\frac{\sqrt{pp'}}{(pp')^{\nu}}.
\label{GS3}
\end{equation}
Then, one can show \cite{Gautier:2015pca} that the substitution of the expression \eqref{GS3} into the formula \eqref{GS1} gives the following expression for the case of 
the large spacelike interval between the points $x$ and $x'$:
\begin{equation}
\frac{1}{(4\pi)^{d/2}}\frac{1}{\Gamma(d/2)}\frac{(\eta\eta')^{\varepsilon}}{\varepsilon|{\bf X}-{\bf X'}|^{2\varepsilon}},
\label{GS4}
\end{equation}
where the parameter $\varepsilon$ is connected  with the $d,\nu$ and the scalar field mass $m$ as follows:
\begin{equation}
\nu = \sqrt{\frac{d^2}{4}-m^2} = \frac{d}{2}-\varepsilon,\ \ \varepsilon \approx \frac{m^2}{d}.
\label{GS5}
\end{equation}
Using the de Sitter invariance and the formula \eqref{GS5}, we can rewrite the formula \eqref{GS4}
as follows:
\begin{equation}
\frac{1}{(4\pi)^{d/2}}\frac{d(2|Z|)^{-\varepsilon}}{m^2\Gamma(d/2)},
\label{GS6}
\end{equation}
where $Z$ is, as usual, the de Sitter invariant function \eqref{z1}.

The symmetrized two-point function calculated in \cite{Gat-Ser} is given by (see Eq.~(46) of \cite{Gat-Ser})
\begin{equation}
\hat{F}(p,p') = \sqrt{pp'}\tilde{F}_{\nu}\left(\frac{c_+}{(pp')^{\bar{\nu}_+-\varepsilon}}+\frac{c_-}{(pp')^{\bar{\nu}_--\varepsilon}}\right).
\label{GS7}
\end{equation} 
Correspondingly, its coordinate representation $F(x,x')$ is related to \eqref{GS7} in the same way that \eqref{GS6} is related to \eqref{GS3}: 
\begin{equation}
F(x,x') = \frac{d}{(4\pi)^{d/2}\Gamma(d/2)}\tilde{F}_{\nu}
\left(
\frac{c_+}{m_+^2}|Z|^{-\varepsilon_+}
+\frac{c_m}{m_-^2}|Z|^{-\varepsilon_-}\right),
\label{GS8}
\end{equation}
where 
\begin{equation}
\tilde{F}_{\nu} = \frac{[2^{\nu}\Gamma(\nu)]^2}{4\pi}.
\label{F-tilde}
\end{equation}
Here the exponents $\varepsilon_+$ and $\varepsilon_-$ are defined as 
\begin{equation}
\varepsilon_{\pm} = \frac{d}{2}-\bar{\nu}_{\pm}+\varepsilon,
\label{GS81}
\end{equation}
where 
\begin{equation}
\bar{\nu}_{\pm}=\sqrt{\nu^2 \pm 2\varepsilon\tilde{\nu}+\varepsilon^2},
\label{GS9}
\end{equation}  
\begin{equation}
\tilde{\nu} = \sqrt{\nu^2+\frac{\sigma_{\rho}}{4\varepsilon^2}},
\label{GS10}
\end{equation}
\begin{equation}
\sigma_{\rho} = \frac{\lambda^2}{\varepsilon^2}\frac{N+2}{24N^2}\frac{\Gamma^2(d/2)}{4\pi^{d+2}},
\label{GS11}
\end{equation}
where $N$ is the number of scalar fields.
The quantities $\nu$ and $\varepsilon$ in Eqs.~\eqref{GS8}-\eqref{GS11} are defined as 
\begin{equation}
\nu = \sqrt{\frac{d^2}{4}-M^2} = \frac{d}{2}-\varepsilon,\ \varepsilon \approx \frac{M^2}{d},
\label{GS12}
\end{equation}
where the mass $M$ is obtained by taking into account the local contribution to the self-energy \cite{Gat-Ser}. The relation between the mass $M$ and the mass parameter $m$ of the Lagrangian is given by the equation 
\begin{equation}
M^2\approx m^2 +\frac{c_d\lambda}{M^2},
\label{GS13}
\end{equation}
where 
\begin{equation}
c_d =\frac{(N+2)d\Gamma\left(\frac{d}{2}\right)}{12 N\cdot 2\pi^{\frac{d}{2}+1}}. 
\label{GS14}
\end{equation}
Solving Eq. \eqref{GS13} perturbatively up to quadratic terms in $\lambda$, one finds
\begin{equation}
M^2=\frac{m^2}{2}+\sqrt{\frac{m^4}{4}+c_d\lambda} = m^2\left(1+\frac{c_d\lambda}{m^4}-\frac{c_d^2\lambda^2}{m^8}\right).
\label{GS15}
\end{equation}

Now we are in a position to calculate the exponents $\varepsilon_{\pm}$. First of all, from Eq. \eqref{GS10} it follows that up to quadratic terms in $\lambda$,
\begin{equation}
\tilde{\nu} = \nu +\frac{\sigma_{\rho}}{8\varepsilon^2\nu}.
\label{GS16}
\end{equation}
Then, from Eq. \eqref{GS9}
it follows that 
\begin{equation}
\bar{\nu}_{\pm} = \nu \pm \varepsilon \pm \frac{\sigma_{\rho}}{8\varepsilon\nu(\nu\pm\varepsilon)},
\label{GS17}
\end{equation}
and hence, 
\begin{equation}
\varepsilon_{+} = \varepsilon-\frac{\sigma_{\rho}}{8\varepsilon\nu^2},
\label{GS18}
\end{equation}
\begin{equation}
\varepsilon_-=3\varepsilon + \frac{\sigma_{\rho}}{8\varepsilon\nu^2}.
\label{GS19}
\end{equation}
Using the formulas written above we obtain the following expression for $\varepsilon_+$:
\begin{eqnarray}
\varepsilon_+ = \frac{m^2}{d}\bigg(1&+&\frac{\lambda d (N+2)\Gamma(d/2)}{24N\pi^{d/2+1}m^4}\nonumber\\&-&\frac{\lambda^2d^2(N+2)(N+5)\Gamma^2(d/2)}{576N^2\pi^{d+2}m^8}\bigg).
\label{GS20}
\end{eqnarray}
Substituting the values $d=3$ and $N=1$ into Eq. \eqref{GS20} and taking into account that the coupling constant $\lambda$ in \cite{Gat-Ser} is 6 times greater than the coupling constant in the present paper, we can see that $\varepsilon_+$ coincides with the exponent of the first term in our Eqs.~\eqref{expt}--\eqref{expz}.

The expression for $\varepsilon_-$ also coincides with the corresponding exponent in our Eqs.~\eqref{expt}--\eqref{expz}, where its is calculated only at $\lambda^0$ order, because the coefficient in front of the exponential is already proportional to $\lambda^2$. 

Let us now calculate the coefficients in front of the exponentials in Eq.~\eqref{GS8}. The coefficients $c_{\pm}$ are given by \cite{Gat-Ser}
\begin{equation}
c_{\pm} = \frac{\tilde{\nu}\pm \nu}{2\tilde{\nu}}.
\label{GS21}
\end{equation}
Using Eqs.~\eqref{GS10} and \eqref{GS11}, we obtain
\begin{equation}
c_+ = 1- \frac{\lambda^2d^2(N+2)\Gamma^2(d/2)}{384N^2\pi^{d+2}m^8}.
\label{GS22}
\end{equation}
\begin{equation}
c_-=\frac{\lambda^2d^2(N+2)\Gamma^2(d/2)}{384N^2\pi^{d+2}m^8}. 
\label{GS23}
\end{equation}
We should also find  the ``masses'' $m_{\pm}$ arising in Eq.~\eqref{GS8}. They are given by
\begin{equation}
\bar{\nu}_{\pm} - \varepsilon = \sqrt{\frac{d^2}{4}-m^2_{\pm}}.
\label{GS24}
\end{equation}
Using the definitions given above, we find that 
\begin{equation}
m_+^2 = M^2 -\frac{\lambda^2 d^2(N+2)\Gamma^2(d/2)}{192N^2\pi^{d+2}}
\label{GS25}
\end{equation}
and 
\begin{equation}
m_-^2=3M^2+\frac{\lambda^2 d^2(N+2)\Gamma^2(d/2)}{192N^2\pi^{d+2}}.
\label{GS26}
\end{equation}
Recall that in Eq.~\eqref{GS8} the expressions $\frac{c_+}{m_+^2}$ and $\frac{c_-}{m_-^2}$ are multiplied by the factor $\frac{d}{(4\pi)^{d/2}\Gamma(d/2)}\tilde{F}_{\nu}$. Hence, the coefficient in front of the second exponential equals
\begin{equation}
\frac{d}{(4\pi)^{d/2}\Gamma(d/2)}
\tilde{F}_{\nu}
\frac{c_-}{m_-^2}
=\frac{\lambda^2d^3(N+2)\Gamma^3(d/2)}{4608N^2\pi^{\frac32d+3}m^{10}}.
\label{GS27}
\end{equation}
Rescaling the coupling constant and substituting $N=1$ and $d=3$ into Eq.~\eqref{GS27}, we see that it is in agreement with the corresponding coefficient in our Eqs.~\eqref{expt}--\eqref{expz}.

The coefficient in front of the first exponential in Eq.~\eqref{GS8} equals
\begin{eqnarray}
\frac{d}{(4\pi)^{d/2}\Gamma(d/2)}\tilde{F}_{\nu}\frac{c_+}{m_+^2}&=&\frac{d\Gamma(d/2)}{4\pi^{\frac{d}{2}+1}m^2}
\bigg(1-\frac{\lambda d(N+2)\Gamma(d/2)}{24N\pi^{\frac{d}{2}+1}m^4}
\nonumber\\&+&\frac{\lambda^2d^2(N+2)(4N+11)\Gamma^2(d/2)}{1152N^2\pi^{d+2}m^8}\bigg).\nonumber \\
\label{GS28}
\end{eqnarray}
After suitable substitutions, the expression \eqref{GS28} coincides with the coefficient of the first term of Eqs.~\eqref{expt}--\eqref{expz}.

\end{document}